\documentclass[aip,jcp,preprint,graphicx]{revtex4-1}
\usepackage{graphicx}
\usepackage[english]{babel}
\usepackage{amsmath}
\usepackage{amssymb}
\usepackage{color}
\usepackage{comment}
\usepackage{microtype}

\definecolor{dgreen}{rgb}{0,.5,0}
\definecolor{dred}{rgb}{.7,.0,.0}

\DeclareMathOperator*{\argmin}{arg\,min}

\begin{document}


\title{Alternative separation of exchange and correlation energies in multi-configuration range-separated density-functional theory} 



\author{Alexandrina Stoyanova}
\affiliation{Laboratoire de Chimie Quantique, Institut de Chimie, CNRS / Universit\'{e} de Strasbourg, 1 rue Blaise Pascal, 67000 Strasbourg, France}
\affiliation{Max-Planck-Institut f\"ur Physik komplexer Systeme, N\"othnitzer Strasse 38, 01187 Dresden, Germany}

\author{Andrew~M.~Teale}
\affiliation{School of Chemistry, University of Nottingham, University Park, Nottingham NG7 2RD, UK}
\affiliation{Centre for Theoretical and Computational Chemistry, Department of Chemistry, University of Oslo, P.O. Box 1033 Blindern, N-0315 Oslo, Norway}

\author{Julien Toulouse} 
\affiliation{Laboratoire de Chimie Th\'eorique, Universit\'e Pierre et Marie Curie and CNRS, 4 place Jussieu, 75005 Paris, France}

\author{Trygve Helgaker}
\affiliation{Centre for Theoretical and Computational Chemistry,
Department of Chemistry, University of Oslo,
P.O. Box 1033 Blindern, N-0315 Oslo, Norway}

\author{Emmanuel Fromager}
\affiliation{Laboratoire de Chimie Quantique, Institut de Chimie, CNRS / Universit\'{e} de Strasbourg, 1 rue Blaise Pascal, 67000 Strasbourg, France}


\date{\today}

\begin{abstract}
The alternative separation of exchange and correlation energies proposed by Toulouse {\it et al.} [Theor.\ Chem.\  Acc.\ {\bf 114}, 305 (2005)] is 
explored in the context of multi-configuration range-separated density-functional theory.  
The new decomposition of the short-range exchange--correlation energy
relies on the auxiliary long-range interacting wavefunction rather than
the Kohn--Sham (KS) determinant. The advantage, relative to the traditional
KS decomposition, is that the wavefunction part of the energy is now
computed with the regular (fully-interacting) Hamiltonian. 
One potential drawback is that, because of double counting, the
wavefunction used to compute the energy cannot be obtained
by minimizing the energy expression with respect to
the wavefunction parameters. The problem is overcome by    
using short-range optimized effective potentials (OEPs). The resulting combination of OEP techniques with wavefunction
theory has been investigated in this work, at the {\it Hartree-Fock}
(HF) and {\it multi-configuration self-consistent-field} (MCSCF)
levels. In the HF case, an analytical expression for the energy
gradient has been derived and implemented. Calculations have been
performed within the short-range local density approximation on H$_2$, N$_2$, Li$_2$ and
H$_2$O. Significant improvements in binding energies are obtained with
the new decomposition of the short-range energy. The importance of
optimizing the short-range OEP at the MCSCF level when static correlation
becomes significant has also been demonstrated for
H$_2$, using a finite-difference gradient.  
The implementation of the analytical gradient for MCSCF wavefunctions is currently in
\end{abstract}


\maketitle 

\section{Introduction}\label{INTRO}

The simultaneous description of dynamical and non-dynamical (static or
strong) electron correlation in atomic and molecular systems, using
low-cost methodologies, remains a challenge for electronic-structure
theory. In particular, standard Kohn--Sham density-functional
theory~\cite{kstheo} (KS-DFT) approximations have enjoyed success in
treating phenomena where the description of short-range dynamical
correlation is paramount but they have been unable 
to provide reliable results whenever static correlation is important---including the description of 
bond breaking, transition-metal compounds, conjugated polymers, and magnetic materials.

Over the years, a great deal of effort has been devoted to understand the shortcomings of density-functional approximations (DFAs) for the treatment of static correlation and to develop new approximations to address this situation. 
Within the framework of KS-DFT, the pragmatic {\it unrestricted}
approach can of course be used for describing bond breaking for example
but, in this
case, space and spin symmetries are broken which is fundamentally not
satisfactory. Recent progress towards DFAs capable of treating static
correlation has been made, for example, by Malet and Gori-Giorgi~\cite{Gori-Giorgi_PRL_2012} and by Becke~\cite{dft-becke_JCP2013}. Other authors have focussed on ensemble-DFT (E-DFT). The utility of E-DFT has been analyzed by Schipper \emph{et al.}~\cite{ensemble_Baerends_1998} and recent implementations of E-DFT variants have been made by Filatov \emph{et al.}~\cite{REKS-Filatov-CPL_1999}, by Chai~\cite{dft-Chai_JCP2012}, and by Nygaard and Olsen~\cite{can-ensemble-Olsen_2013}. 

Beyond the framework of standard KS-DFT, a number of groups have pursued
the idea of hybridizing density-functional and {\it multi-configuration
self-consistent-field} (MCSCF) approaches. The goal of these approaches is to treat static correlation using the flexibility of the MCSCF expansion, 
whilst treating dynamical correlation using density functionals \cite{ColleandSalvetti_JCP_1990, savinijqc, savinijqc2, MiehlichStollSavin, DoubleCountothers1, Stoll2003141, Shaik_JChemTheoryComputation, Malcolm1998121, JMcdouall_MP_2003, DoubleCountothers1b, DoubleCountothers2, NakataUkaietal_IJQC_2006, YamanakaNakataetal_IJQC_2006, UkaiNakataetal_MP_2007,PhysRevA.75.012503}. A number of MCSCF-DFT hybrid approaches have been proposed, 
including the complete-active-space-DFT (CAS-DFT) schemes of Gr\"affenstein \emph{et al.} \cite{DoubleCountothers1}, 
Gusarov \emph{et al.} \cite{DoubleCountothers2}, and Miehlich \emph{et
al.} \cite{MiehlichStollSavin} A multi-configuration extension of
KS-DFT using {\it optimized effective potential} techniques has also been proposed by Weimer {\it et
al.}~\cite{JCP08_Gorling_mcoep} A challenge for these methodologies is
to avoid double counting of correlation effects, as the density
functionals utilized in these approaches depend on the MCSCF expansions.
Tackling the double counting problem is a difficult task as illustrated
by Kurzweil \textit{et al.} \cite{MCDFT_Head-Gordon_MolPhys2009} in
their analysis on the mapping of interacting onto partially interacting
system. 

A key step in avoiding the doubling counting problems of the MCSCF-DFT
hybrid approaches was the proposal of Savin \emph{et
al.}~\cite{savinbook, stollsavinbook} to divide the Coulomb interaction
into long-range (lr) and short-range (sr) contributions. The
introduction of this range-separated approach has led to a wide variety
of hybrid wave-function/DFT approaches; in the present context, we note
the {\it multi-configuration short-range DFT} (MC-srDFT) approach of
Refs.~\onlinecite{jesperthesis,dft-Fromager-JCP2007a,JCPunivmu2}. 
Range separation has proven to be of great utility for the treatment of dispersion interactions, where the simple physical intuition that interactions are long-ranged can be leveraged to provide a clean division of labour between the density-functional and wave-function contributions.\cite{srdftnancy,ccsrdft,rpa-srdft_toulouse,nevpt2srdft2, rpa-srdft_scuseria, rpa-srdft_toulouse_2011,TealeDisp} 
However, for the description of static correlation, this approach is
less effective since static correlation may not be interpreted as
predominantly long-ranged. Thus, even for systems such as stretched
H$_2$, where static correlation is expected to be dominant, the
corresponding short-ranged density-functional still plays a significant
role~\cite{AMTOptseplrsr} and errors associated with these
approximations can lead to significant errors in binding energies,
albeit with an improved shape for the potential energy
curve.\cite{dft-Fromager-JCP2007a} 
These considerations are reflected in the fact that a recently proposed MC-DFT approach based on a simple linear (rather than range-dependent) decomposition of the Coulomb interaction delivers similar accuracy in practice.\cite{Sharkas_JCP}

To overcome some of the difficulties associated with the MC-srDFT models for the description of static correlation, 
Toulouse \emph{et al.}~\cite{TousrXmd} proposed an alternative separation of the short-range exchange and correlation energies, which may be more natural in the context of hybrid methodologies that incorporate multi-configurational components. The practical performance of methods using this partitioning is investigated in the present work.

The paper is organized as follows. In Sec.~\ref{THEORY}, exact
and approximate formulations of range-separated DFT are 
presented. The differences between the traditional and the alternative
short-range exchange--correlation energy decompositions are highlighted. 
For the latter, long-range HF 
and MCSCF approximations are combined with short-range OEPs in order to overcome double counting problems. Details of the various range-separated schemes that
have been implemented are given in Sec.~\ref{comp_details_sec}, 
while the results obtained for H$_2$, N$_2$, H$_2$O,
and Li$_2$ are discussed in Sec.~\ref{sec:results}. 
Finally, conclusions are given in Sec.~\ref{conclusion_sec}. 

\section{Theory}\label{THEORY}

In this section, we present the theory underlying 
the various multi-configuration range-separated DFT models assessed in
Sec.~\ref{sec:results}. We first introduce in
Sec.~\ref{subsec:srdft} the exact multi-determinantal
extension of KS-DFT based 
on range separation. The conventional KS decomposition of the
short-range exchange--correlation density functional, as well as an alternative one 
that relies on the long-range interacting wavefunction (rather than the
KS determinant), are next discussed in
Secs.~\ref{subsec:sep_srxc} and \ref{subsec:alternative_sep_srxc}. The
combination of OEPs with multi-determinant wavefunctions and density-functionals is then considered
in Sec.~\ref{susubsec:Exact-oep-dft-wft}. Models based on HF and MC
wavefunctions are discussed in Sec.~\ref{subsubsec:approx-sroep}.
The derivation of the analytical energy gradient for the OEP optimization
at the HF level is
finally presented in Sec.~\ref{subsecHFOEP}. A summary is given
in Sec.~\ref{subsec:summary}. 

\subsection{Multi-determinant range-separated DFT}\label{subsec:srdft}
In multi-determinant range-separated DFT,\cite{savinbook, SavinDFT,stollsavinbook} 
referred to as \emph{short-range DFT} (srDFT) in the following, the regular
two-electron Coulomb interaction is split into long- and short-range parts,
\begin{align}\label{lrsrsep}
w_{\rm ee}(r_{12})&=1/r_{12} =w^{\rm lr,\mu}_{\rm ee}(r_{12})+w^{\rm sr,\mu}_{\rm ee}(r_{12}),
\end{align}
where $\mu$ is a parameter that controls the range separation. In this
work, the long-range interaction is based on the error function 
\begin{align}\label{lrerf} 
{w^{\rm lr,\mu}_{\rm ee}(r_{12}) =\frac{{\rm erf}(\mu r_{12})}{r_{12}}.} 
\end{align}
The universal Levy--Lieb functional~\cite{LevyF,LiebF} 
\begin{align}\label{univfunc}
F[n] &= \underset{\Psi\rightarrow n}{\rm min}\langle \Psi\vert \hat{T}+\hat{W}_{\rm ee}\vert\Psi\rangle,
\end{align}
where $\hat{T}$ is the kinetic energy operator and $\hat{W}_{\rm ee}$
the two-electron repulsion operator,
can then be rewritten as 
\begin{align}\label{splitunivfunc}
F[n] &= F^{\rm lr,\mu}[n] + E^{\rm sr,\mu}_{\rm Hxc}[n],
\end{align}
with the universal long-range functional defined as 
\begin{align}\label{lrunivfunc}
F^{\rm lr,\mu}[n] & = \underset{\Psi\rightarrow n}{\rm min}\langle \Psi\vert \hat{T}+\hat{W}^{\rm lr, \mu}_{\rm ee}\vert\Psi\rangle \nonumber \\
& = \langle\Psi^{\mu}[n]\vert \hat{T}+\hat{W}^{\rm lr,\mu}_{\rm ee}\vert\Psi^{\mu}[n]\rangle.
\end{align}
The minimizing wavefunction $\Psi^{\mu}[n]$ in Eq.~(\ref{lrunivfunc})
corresponds to the ground state of the auxiliary long-range interacting
system with density $n$. When connecting the auxiliary and physical systems
by a generalized adiabatic-connection path,\cite{Yang:1998p441,AMTOptseplrsr,SavRev} the complementary
\emph{short-range Hartree--exchange--correlation} (srHxc) energy can be expressed as 
\begin{align}\label{ACsrHxcfun}
E ^{\rm sr,\mu}_{\rm Hxc}[n] &= \int^1_{\mu/(1+\mu)} \mathcal{W}^{\nu}_{\rm Hxc}[n]\;\mathrm{d}\nu,
\end{align}
where the srHxc integrand is given by 
\begin{align}\label{srHxcintegrand}
\mathcal{W}^{\nu}_{\rm Hxc}[n] &= \left\langle\frac{\partial \hat{W}^{\rm lr,\nu /(1-\nu)}_{\rm ee}}{\partial \nu}\right\rangle_{\Psi^{\nu/(1-\nu)}[n]}.
\end{align}
Note that the expression in Eq.~(\ref{ACsrHxcfun}) relies on the density constraint
\begin{align}\label{dens_constraint}
n_{\Psi^{\nu /(1-\nu)}[n]}({\bf r}) &=n({\bf r}),\hspace{0.5cm} 0\leq\nu\leq1.
\end{align}
According to the variational principle~\cite{hktheo} and Eq.~(\ref{splitunivfunc}), the exact expression for the ground-state energy
of an electronic system becomes 
\begin{align}\label{Varprinciple}
E &= \underset{n}{\rm min}\left\{ F[n]+\int {\rm d}{\bf r}\,v_{\rm ne}({\bf r})\,n({\bf r}) \right\} \nonumber \\
&= \underset{n}{\rm min}\left\{\langle \Psi^{\mu}[n] \vert \hat{T}+\hat{W}^{\rm lr,\mu}_{\rm ee}+\hat{V}_{\rm ne}\vert \Psi^{\mu}[n]\rangle +E^{\rm sr,\mu}_{\rm Hxc}[n]\right\},
\end{align}
where $\hat{V}_{\rm
ne}=\int {\rm d}{\bf r}\,v_{\rm ne}({\bf r})\,\hat{n}({\bf r})$ 
is the nuclear
potential operator. The exact ground-state energy can also
be obtained by a minimization over wavefunctions
\begin{align}\label{rsDFTgs}
E&=\underset{\Psi}{\rm min}\left\lbrace \langle \Psi \vert \hat{T}+\hat{W}^{\rm lr,\mu}_{\rm ee}+\hat{V}_{\rm ne}\vert \Psi \rangle+E^{\rm sr,\mu}_{\rm Hxc}[n_{\Psi}] \right\rbrace \nonumber \\
 &=\langle \Psi^{\mu} \vert \hat{T}+\hat{W}^{\rm lr,\mu}_{\rm ee}+\hat{V}_{\rm ne}\vert \Psi^{\mu} \rangle+E^{\rm sr,\mu}_{\rm Hxc}[n_{\Psi^{\mu}}],
\end{align}
where the minimizing wavefunction $\Psi^{\mu}$ fulfills the self-consistent equation 
\begin{align}\label{fictselfconseq}
(\hat{T}+\hat{W}^{\rm lr,\mu}_{\rm ee}+\hat{V}^{\mu}[n_{\Psi^{\mu}}])\vert\Psi^{\mu}\rangle=\mathcal{E}^{\mu} \vert\Psi^{\mu}\rangle,
\end{align}
with
\begin{align}\label{vsrmu}
\hat{V}^{\mu}[n]=\int {\rm d}{\bf r}\,\left(v_{\rm ne}({\bf r})+ \frac{\delta E^{\rm sr,\mu}_{\rm Hxc}}{\delta n ({\bf r})}[n] \right)\hat{n}({\bf r}). 
\end{align}
While $\hat{W}^{\rm lr,\mu}_{\rm ee}$ vanishes and
Eq.~(\ref{fictselfconseq}) reduces to the conventional KS equation at $\mu=0$, the
full Schr\"odinger equation is recovered in the $\mu \rightarrow +\infty$
limit, as $\hat{W}^{\rm lr,\mu}_{\rm ee}$ reduces to the regular
two-electron repulsion and the short-range interaction vanishes.
For intermediate $\mu$ values, $0<\mu<+\infty$, a hybrid wave-function/DFT description
is obtained. Then, in contrast to traditional KS-DFT, the exact auxiliary wavefunction 
$\Psi^{\mu}$ is generally 
multi-determinantal owing to the explicit description of long-range
interactions. 

The srDFT approximation obtained by restricting the
minimization in Eq.~(\ref{rsDFTgs}) to single determinants is in the following referred
to as HF-srDFT; this approximation was referred to as range-separated hybrid (RSH) theory in Ref.~\onlinecite{srdftnancy}.
To describe multi-configurational electronic systems, a long-range MCSCF description has 
also been proposed,\cite{jesperthesis,dft-Fromager-JCP2007a} leading to the MC-srDFT model. In this work, we consider
both schemes. 

\subsection{KS decomposition of the short-range energy}\label{subsec:sep_srxc}

The conventional decomposition of the srHxc energy is analogous to that in standard KS-DFT~\cite{erferfgaufunc}, 
\begin{align} \label{srDFTfunhxcdef}
E^{\rm sr,\mu}_{\rm Hxc}[n]&=E^{\rm sr,\mu}_{\rm H}[n]+E^{\rm sr,\mu}_{\rm x}[n]+E^{\rm sr,\mu}_{\rm c}[n], 
\end{align}
where the short-range Hartree energy is defined as 
\begin{align} \label{srDFTfunHdef}
E^{\rm sr,\mu}_{\rm H}[n]= \frac{1}{2}\int\int {\rm d}{\mathbf{r}}{\rm d}{\mathbf{r'}}n(\mathbf{r})n(\mathbf{r'})w^{\rm sr,\mu}_{\rm ee}\left(\vert {\bf r}-{\bf r'} \vert\right),
\end{align}
and the exact short-range exchange energy is calculated from 
the non-interacting KS determinant $\Phi^{\rm
KS}[n]$ with density $n$. The Hartree--exchange integrand is then
obtained from Eq.~(\ref{srHxcintegrand}) by replacing the long-range 
interacting wavefunction by $\Phi^{\rm KS}[n]$    
\begin{align}\label{srHxintegrand}
\mathcal{W}^{\nu}_{\rm Hx}[n]&=\left\langle\frac{\partial \hat{W}^{\rm lr,\nu /(1-\nu)}_{\rm ee}}{\partial \nu}\right\rangle_{\Phi^{\rm KS}[n]},
\end{align}
which defines the short-range exchange energy, according to
Eq.~(\ref{ACsrHxcfun}), as
\begin{align} \label{srxfundef}
E^{\rm sr,\mu}_{\rm x}[n]&=\int^1_{\mu/(1+\mu)} \mathcal{W}^{\nu}_{\rm Hx}[n]\;\mathrm{d}\nu -E^{\rm sr,\mu}_{\rm H}[n]\\
&=\langle \Phi^{\rm KS}[n]\vert\hat{W}^{\rm sr, \mu}_{\rm ee}\vert\Phi^{\rm KS}[n]\rangle -E^{\rm sr,\mu}_{\rm H}[n]. \nonumber
\end{align}
The corresponding correlation integrand is then given by 
\begin{align}\label{srcintegrand}
\mathcal{W}^{\nu}_{\rm c}[n]
&=\mathcal{W}^{\nu}_{\rm Hxc}[n]-\mathcal{W}^{\nu}_{\rm Hx}[n] \nonumber \\
&=\left\langle\frac{\partial \hat{W}^{\rm lr,\nu /(1-\nu)}_{\rm ee}}{\partial \nu}\right\rangle_{\Psi^{\nu/(1-\nu)}[n]} -\left\langle \frac{\partial \hat{W}^{\rm lr,\nu /(1-\nu)}_{\rm ee}}{\partial \nu}\right\rangle_{\Phi^{\rm KS}[n]},
\end{align}
leading to the following expression for the exact complementary
short-range correlation energy 
\begin{align}\label{ACsrcfun}
E ^{\rm sr,\mu}_{\rm c}[n]&=\int^1_{\mu/(1+\mu)} \mathcal{W}^{\nu}_{\rm c}[n]\;\mathrm{d}\nu.
\end{align}
Note that, according to Eqs.~(\ref{splitunivfunc}),
(\ref{lrunivfunc}) and (\ref{srxfundef}),
this energy can also be expressed
as  
\begin{align} \label{srDFTfuncdef}
E ^{\rm sr,\mu}_{\rm c}[n]=E_{\rm c}[n]&+ \langle \Phi^{\rm KS}[n]\vert\hat{T}+\hat{W}^{\rm lr, \mu}_{\rm ee}|\Phi^{\rm KS}[n]\rangle \nonumber \\
&-\langle \Psi^{\mu}[n] \vert \hat{T}+\hat{W}^{\rm lr,\mu}_{\rm ee}\vert \Psi^{\mu}[n]\rangle,
\end{align}
where $E_{\rm c}[n]$ is the regular correlation density-functional
energy, recovered in the $\mu=0$ limit.
It is clear from Eq.~(\ref{srDFTfuncdef}) that the complementary short-range
correlation density functional contains the purely short-range
correlation effects as well as their coupling with long-range
correlation.\cite{TouSav_srlrcorr}
Various short-range functionals have been developed for
practical srDFT calculations at the local density (srLDA),~\cite{TouSav_srlrcorr, toulda, srmdCref} generalized
gradient,\cite{pbehsea,pbehseb,ccsrdft, erferfgaufunc, ccsrdft2, TouColSavin_JCP2005} and meta-generalized gradient\cite{goll_srmeta-gga} levels of approximation. 
These functionals have been successfully employed with a number of post-HF-srDFT and post-MC-srDFT long-range correlation treatments to describe dispersion.~\cite{srdftnancy,ccsrdft,rpa-srdft_toulouse,nevpt2srdft2, rpa-srdft_scuseria, rpa-srdft_toulouse_2011} However,
for systems with significant static correlation, they are usually not accurate enough.\cite{dft-Fromager-JCP2007a,JCPunivmu2}
To improve upon the description of the short-range energy, we consider in the following an
alternative separation of exchange and correlation energies.
 
\subsection{Alternative decomposition of the short-range energy}\label{subsec:alternative_sep_srxc}

As pointed out by Toulouse, Gori-Giorgi, and Savin,\cite{TousrXmd, PaolasrXmd} 
it is more natural, in the context of srDFT, to define the short-range exchange energy in terms of the
multi-determinantal wavefunction $\Psi^{\mu}[n]$ introduced in
Eq.~(\ref{lrunivfunc}). This observation leads to the  
following decomposition of the srHxc
density-functional energy 
\begin{align}\label{Ehxc_md}
E^{\rm sr,\mu}_{\rm Hxc}[n]=E^{\rm sr,\mu}_{\rm H}[n]+E^{\rm sr,\mu}_{\rm x, md}[n] +E^{\rm sr,\mu}_{\rm c,md}[n], 
\end{align} 
where, what is referred to as the short-range multideterminantal (MD) 
exchange functional in Ref.~\onlinecite{PaolasrXmd}, is defined as
\begin{align}\label{srmdEXXfunc}
E^{\rm sr,\mu}_{\rm x, md}[n]= \langle \Psi^{\mu}[n]\vert \hat{W}^{\rm sr,\mu}_{\rm ee}\vert\Psi^{\mu}[n]\rangle -E^{\rm sr,\mu}_{\rm H}[n].
\end{align} 
This expression arises naturally from Eqs.~(\ref{ACsrHxcfun}) and~(\ref{srHxcintegrand})
when replacing the $\nu$-dependent wavefunction in the integrand by the one obtained
with the lower integration limit $\nu=\mu/(1+\mu)$, namely
$\Psi^{\mu}[n]$. We thus define the MD
Hartree--exchange integrand as
\begin{align}\label{srmdHxintegrand}
\mathcal{W}^{\mu,\nu}_{\rm Hx, md}[n] &= \left\langle\frac{\partial \hat{W}^{\rm lr,\nu /(1-\nu)}_{\rm ee}}{\partial \nu}\right\rangle_{\Psi^{\mu}[n]}.
\end{align}
The short-range MD exchange energy is then obtained as follows, according to
Eq.~(\ref{ACsrHxcfun}), 
\begin{align} \label{srmdxfundef}
E^{\rm sr,\mu}_{\rm x, md}[n]&=\int^1_{\mu/(1+\mu)} \mathcal{W}^{\mu,\nu}_{\rm Hx, md}[n]\;\mathrm{d}\nu -E^{\rm sr,\mu}_{\rm H}[n],
\end{align}
leading to Eq.~(\ref{srmdEXXfunc}).
We emphasize that, for $\mu>0$, $\Psi^{\mu}[n]$ differs from
the KS determinant. As a result, the expression for the ``exchange" energy 
in Eq.~(\ref{srmdEXXfunc}) contains a correlation 
contribution, in addition to the
short-range exchange energy. 
Note also that the complementary short-range correlation functional in
Eq.~(\ref{Ehxc_md}) differs from the conventional one introduced in
Eq.~(\ref{srDFTfunhxcdef}):
\begin{align} \label{srcCfun}
E ^{\rm sr,\mu}_{\rm c,md}[n]&=E ^{\rm sr,\mu}_{\rm c}[n]+ \langle \Phi^{\rm KS}[n]\vert\hat{W}^{\rm sr, \mu}_{\rm ee}\vert\Phi^{\rm KS}[n]\rangle \nonumber\\
&-\langle \Psi^{\mu}[n] \vert \hat{W}^{\rm sr,\mu}_{\rm ee}\vert \Psi^{\mu}[n]\rangle.
\end{align}
This also becomes clear when expressing the MD short-range correlation energy,
\begin{align} \label{srmdcfundef}
E^{\rm sr,\mu}_{\rm c, md}[n]&= \int^1_{\mu/(1+\mu)} \mathcal{W}^{\mu,\nu}_{\rm c, md}[n]\;\mathrm{d}\nu,
\end{align}
in terms of the corresponding correlation integrand
\begin{align}\label{srmdcintegrand}
\mathcal{W}^{\mu,\nu}_{\rm c, md}[n]
&=\mathcal{W}^{\nu}_{\rm Hxc}[n]-\mathcal{W}^{\mu,\nu}_{\rm Hx, md}[n] \nonumber \\
&=\left\langle\frac{\partial \hat{W}^{\rm lr,\nu /(1-\nu)}_{\rm ee}}{\partial \nu}\right\rangle_{\Psi^{\nu/(1-\nu)}[n]} - \left\langle \frac{\partial \hat{W}^{\rm lr,\nu /(1-\nu)}_{\rm ee}}{\partial \nu}\right\rangle_{\Psi^{\mu}[n]},
\end{align}
which differs from the conventional integrand in Eq.~(\ref{srcintegrand}) only in the use of
$\Psi^{\mu}[n]$ rather than $\Phi^\text{KS}[n]$ in the last (subtracted) term.
Note that the correlation energies $E_\text c[n]$ and $E ^{\rm sr,\mu}_{\rm c,md}[n]$ are related in a simple manner,
as seen by
inserting Eq.~(\ref{srDFTfuncdef}) into Eq.~(\ref{srcCfun}) and rearranging,
\begin{align} \label{srcCfunX}
&\langle \Psi^{\mu}[n] \vert \hat T + \hat{W}_{\rm ee}\vert \Psi^{\mu}[n]\rangle + E ^{\rm sr,\mu}_{\rm c,md}[n] = \langle \Phi^{\rm KS}[n]\vert \hat T + \hat{W}_{\rm ee}\vert\Phi^{\rm KS}[n]\rangle + E_{\rm c}[n],
\end{align}
where the long-range correlation and its coupling with the short-range
interaction is contained in the expectation value on the left-hand side but in the correlation functional on the right-hand side. 
Substitution of the srHxc energy decomposition in Eqs.~(\ref{Ehxc_md})
and~(\ref{srmdEXXfunc}) back into the ground-state energy expression in Eq.~(\ref{Varprinciple})
leads to
\begin{align}\label{energymin2}
E=\underset{n}{\rm min} \left\{ \langle \Psi^{\mu}[n] \vert \hat{T}+\hat{W}_{\rm ee}+\hat{V}_{\rm ne}\vert\Psi^{\mu}[n]\rangle +E^{\rm sr,\mu}_{\rm c, md}[n] \right\}. 
\end{align}
Since $\Psi^{\mu}[n_{\Psi^{\mu}}]$=$\Psi^{\mu}$, we conclude from
Eq.~(\ref{rsDFTgs})
that the exact ground-state energy can be re-expressed as
\begin{align}\label{energymin3}
E&=\langle \Psi^{\mu}\vert \hat{T}+\hat{W}_{\rm ee}+\hat{V}_{\rm ne}\vert\Psi^{\mu}\rangle +E^{\rm sr,\mu}_{\rm c, md}[n_{\Psi^{\mu}}].
\end{align}
We emphasize that the expression in Eq.~(\ref{energymin3}) is exact when the energy is calculated from the self-consistent
wavefunction in Eq.~(\ref{rsDFTgs}). 
We here introduce an approximation, 
the {\it range-separated hybrid} model with {\it full}-range integrals (RSHf),
where the energy in Eq.~(\ref{energymin3}) is instead computed from the HF-srDFT wavefunction.
A multi-configuration extension is obtained when using the
MC-srDFT rather than HF-srDFT wavefunction, defining thus a {\it range-separated
multi-configuration hybrid} model
with {\it full}-range integrals (RSMCHf).

In the RSHf and RSMCHf schemes, the
wavefunctions are optimized with the conventional
short-range exchange--correlation density-functional, while the energy
is computed with the alternative separation of exchange and
correlation energies. As discussed in Sec.~\ref{sec:results}, this approach 
may not be sufficiently accurate when approximate functionals are
used, especially when static correlation becomes important.
The alternative decomposition of the short-range energy should then be
used for the optimization of the wavefunction. However, unlike in srDFT, the minimization over
densities in Eq.~(\ref{energymin2}) cannot be replaced by a minimization over wavefunctions,  
\begin{align}\label{CAS-DFT-like}
E\neq\underset{\Psi}{\rm min}\left\lbrace\langle \Psi \vert\hat{T}+\hat{W}_{\rm ee}+\hat{V}_{\rm ne}\vert\Psi\rangle +E^{\rm sr,\mu}_{\rm c, md}[ n_{\Psi}]\right\rbrace,
\end{align}
simply because the minimizing wavefunction would be the ground state of
a fully-interacting system, leading thus to double counting. On the other hand, invoking
the one-to-one correspondence between densities and local potentials, a
multi-determinant extension of KS-OEP schemes can be formulated. 

\subsection{Multi-determinant range-separated OEP approach}

\subsubsection{Exact formulation}\label{susubsec:Exact-oep-dft-wft}

When using the local potential rather than the density as a basic variable,
the exact ground-state energy expression in Eq.~(\ref{energymin2}) can
be rewritten as~\cite{TousrXmd}
\begin{align}\label{Gsenevmu}
E=&\underset{v}{\rm min} \left\{ \langle \Psi^{\mu}[v] \vert \hat{T}+\hat{W}_{\rm ee}+\hat{V}_{\rm ne}\vert\Psi^{\mu}[v]\rangle \right. + \left. E^{\rm sr,\mu}_{\rm c, md}[n_{\Psi^{\mu}[v]}] \right\}, 
\end{align}
where $\Psi^{\mu}[v]$ is the ground state of the
long-range interacting Hamiltonian with the local potential $v$: 
\begin{align}\label{psimuvfrommin}
\Psi^{\mu}[v] =\!\argmin\limits_\Psi \!\left\{\!\langle \Psi \vert \hat{T}+\hat{W}^{\rm lr,\mu}_{\rm ee}\vert\Psi\rangle + \!\! \int \!\! {\rm d}{\mathbf r}\,v({\mathbf r}){n}_{\Psi}({\bf r})\! \right\}.
\end{align}
Since $\Psi^{\mu}[v]$ is the solution to a (linear) eigenvalue
equation, the formulation in Eqs.~(\ref{Gsenevmu}) and
(\ref{psimuvfrommin}) will be
referred to as {\it non-self-consistent}. According to Eq.~(\ref{srmdcfundef}),
the MD short-range
correlation energy vanishes as $\mu\rightarrow+\infty$
and the minimizing potential in
Eq.~(\ref{Gsenevmu}) is simply the nuclear potential. Regular wave-function theory
is then recovered. On the other hand, when $\mu=0$,
the energy in Eq.~(\ref{Gsenevmu}) reduces to a KS-OEP energy where the exact-exchange (EXX)
term~\cite{Yang_prl_oep} is used in conjunction
with the standard correlation density functional. 

When
$0<\mu<+\infty$, we obtain a rigorous combination of wave-function and KS-OEP density-functional approaches, referred to as srOEP in the following.
According to Eqs.~(\ref{fictselfconseq}), (\ref{vsrmu}) and (\ref{energymin3}), the exact 
minimizing potential is then given by 
\begin{align}\label{exactvmu1stformulation}
v^{\mu}({\bf r})=v_{\rm ne}({\bf r})+ \frac{\delta E^{\rm sr,\mu}_{\rm Hxc}}{\delta n ({\bf r})}[n_{\Psi^{\mu}}],
\end{align}   
which can be rewritten as 
\begin{align}\label{exactvmu1stformulation2}
v^{\mu}({\bf r})=v_{\rm ne}({\bf r}) &+ \frac{\delta E^{\rm sr,\mu}_{\rm H}}{\delta n ({\bf r})}[n_{\Psi^{\mu}}] + \frac{\delta E^{\rm sr,\mu}_{\rm x, md}}{\delta n ({\bf r})}[n_{\Psi^{\mu}}] \nonumber\\
&+ \frac{\delta E^{\rm sr,\mu}_{\rm c, md}}{\delta n ({\bf r})}[n_{\Psi^{\mu}}],
\end{align}
in terms of the srHxc decomposition in
Eq.~(\ref{Ehxc_md}). Since the MD short-range correlation energy 
is here an explicit functional of the density, for which
local density approximations have been proposed,\cite{TousrXmd,PaolasrXmd} only the MD short-range exchange part 
 needs to be optimized in Eq.~(\ref{exactvmu1stformulation2}), being
an implicit functional of the density according to
Eq.~(\ref{srmdEXXfunc}).
Note that the srOEP energy in 
Eq.~(\ref{Gsenevmu}) can be rewritten as
\begin{align}\label{Gsenevmutilde}
E &=\underset{v}{\rm min} \left\{ \langle \tilde{\Psi}^{\mu}[v] \vert \hat{T}+\hat{W}_{\rm ee}+\hat{V}_{\rm ne}\vert\tilde{\Psi}^{\mu}[v]\rangle +  E^{\rm sr,\mu}_{\rm c, md}[n_{\tilde{\Psi}^{\mu}[v]}] \right\},
\end{align}
where the auxiliary 
wavefunction $\tilde{\Psi}^{\mu}[v]$ is obtained for a given local
potential $v$ as follows:
\begin{align}\label{psimuvfromminsc}
\tilde{\Psi}^{\mu}[v] = \argmin\limits_\Psi &\left\{ \langle \Psi \vert \hat{T}+\hat{W}^{\rm lr,\mu}_{\rm ee} +\hat{V}_{\rm ne} \vert \Psi\rangle +E^{\rm sr,\mu}_{\rm H}[n_{\Psi}] +E^{\rm sr,\mu}_{\rm c, md}[n_{\Psi}] +\int {\rm d}{\mathbf r}\;v({\mathbf r})\,{n}_{\Psi}({\bf r}) \right\}.
\end{align}
Indeed, as $\tilde{\Psi}^{\mu}[v]$ satisfies the self-consistent equation  
\begin{align}\label{tildepsimuvsceq}
&\left(\hat{T}+\hat{W}^{\rm lr,\mu}_{\rm ee}+\hat{V}_{\rm ne} +\int {\rm d}{\bf r}\, \frac{\delta E^{\rm sr,\mu}_{\rm H}}{\delta n({\bf r})}[n_{\tilde{\Psi}^{\mu}[v]}]\,\hat{n}(\mathbf{r}) \right. \nonumber \\
&\left. +\int {\rm d}{\bf r}\, \left[ \frac{\delta E^{\rm sr,\mu}_{\rm c,md}}{\delta n({\bf r})}[n_{\tilde{\Psi}^{\mu}[v]}] + v({\mathbf r}) \right]\,\hat{n}(\mathbf{r}) \right) \vert \tilde{\Psi}^{\mu}[v]\rangle \nonumber \\
&=\tilde{\mathcal{E}}^{\mu}[v]\vert\tilde{\Psi}^{\mu}[v]\rangle,
\end{align}
it is clear from Eq.~(\ref{exactvmu1stformulation2}) that the minimum in
Eq.~(\ref{Gsenevmutilde}) is reached for the local
potential     
\begin{align}\label{minpottilde}
\tilde{v}^{\mu}({\mathbf r})= \frac{\delta E^{\rm sr,\mu}_{\rm x, md}}{\delta n({\mathbf r})}[n_{\Psi^{\mu}}]
\end{align}
since
$\tilde{\Psi}^{\mu}[\tilde{v}^{\mu}]=\Psi^\mu$. 
The formulation in Eqs.~(\ref{Gsenevmutilde}) and (\ref{psimuvfromminsc}) will be referred
to as {\it self-consistent}. It is equivalent to the non-self-consistent
formulation in Eq.~(\ref{Gsenevmu}) as long as there are no restrictions
in
the form of the optimized potential. This statement holds if an
approximate MD short-range correlation density functional is used in
conjunction with approximate long-range
interacting HF or MCSCF wavefunctions, as considered in the
rest of this work.\\ 
Since optimized potentials are usually expanded in a
finite basis,\cite{Yang_prl_oep} the non-self-consistent and self-consistent formulations in Eqs.~(\ref{Gsenevmu}) and~(\ref{Gsenevmutilde}),
respectively, may give different results if the
basis set in the first formulation is not sufficiently large to describe 
both short-range Hartree and
MD correlation density-functional potentials accurately. The advantage of the
second formulation lies in the fact that the basis set is used only to
represent the MD exchange part of the short-range
potential, see Eq.~(\ref{minpottilde}); the remaining Hartree and MD correlation
short-range contributions are 
calculated as functional derivatives, see Eq.~(\ref{tildepsimuvsceq}). The drawback with respect to the implementation is related
to
the computation of the energy gradient needed for optimizing the
potential. As discussed in
Sec.~\ref{subsecHFOEP}, the gradient requires the calculation of the linear
response function for the
long-range interacting wavefunction.
The self-consistent formulation
is thus less trivial to implement, requiring the implementation of second-order functional derivative
contributions (kernel) to the linear response
equations.\cite{fromager2013} 
All approximate srOEP models introduced in the following are therefore based
on the non-self-consistent formulation in Eq.~(\ref{Gsenevmu}).

\subsubsection{Approximate formulations}\label{subsubsec:approx-sroep}

As in srDFT, the exact auxiliary wavefunction in srOEP is
multi-determinantal, being the ground state of a long-range
interacting system. In the simplest HF-srOEP approach, the minimization 
in Eq.~(\ref{psimuvfrommin}) is over
single-determinantal wavefunctions $\Phi$, 
\begin{align}\label{phimuvfromminhf}
\Phi^{\mu}[v]=\argmin\limits_{\Phi} \left\{\!\langle \Phi \vert \hat{T}+\hat{W}^{\rm lr,\mu}_{\rm ee} \vert\Phi\rangle + \!\! \int \!\! {\rm d}{\mathbf r}\;v({\mathbf r}){n}_{\Phi}({\bf r}) \!\right\}.
\end{align}
The HF-srOEP energy is then defined as 
\begin{align}\label{GsenevmuHF}
E_{\mbox{\tiny HF}}^{\mbox{\tiny srOEP}}=\underset{v}{\rm min} &\left\{ \langle \Phi^{\mu}[v] \vert \hat{T}+\hat{W}_{\rm ee}+\hat{V}_{\rm ne}\vert\Phi^{\mu}[v]\rangle +  E^{\rm sr,\mu}_{\rm c, md}[n_{\Phi^{\mu}[v]}] \right\}.
\end{align}
A multi-configurational extension is obtained by restricting the
minimization in Eq.~(\ref{psimuvfrommin}) to MCSCF wavefunctions that
belong to a given active space $S_M$:
\begin{align}\label{phimuvfromminmc}
\Psi^\mu_M[v]&= \argmin\limits_{\Psi\in S_M} \left\{\langle \Psi \vert \hat{T}+\hat{W}^{\rm lr,\mu}_{\rm ee} \vert\Psi\rangle + \int {\rm d}{\mathbf r}\;v({\mathbf r})\,{n}_{\Psi}({\bf r}) \right\}.
\end{align}
This approach leads to the MC-srOEP energy expression
\begin{align}\label{GsenevmuMC}
E_{\mbox{\tiny MC}}^{\mbox{\tiny srOEP}}=\underset{v}{\rm min} &\left\{ \langle \Psi_M^{\mu}[v] \vert \hat{T}+\hat{W}_{\rm ee}+\hat{V}_{\rm ne}\vert\Psi_M^{\mu}[v]\rangle + E^{\rm sr,\mu}_{\rm c, md}[n_{\Psi_M^{\mu}[v]}] \right\},
\end{align}
similar to the CAS-DFT 
energy expression of Refs.~\onlinecite{DoubleCountothers1,DoubleCountothers2}, based on the
regular Hamiltonian and a complementary correlation functional.
However, unlike in CAS-DFT, the correlation functional in the MC-srOEP method is
universal in the sense that it does not depend on the active space $S_M$. 
In addition,  
the minimization over local potentials (rather than over wavefunctions) ensures that the MCSCF model is applied to a long-range
interacting Hamiltonian. As a result, the active space can be enlarged
to the full configuration-interaction (FCI) limit with no risk of
double counting correlation effects.\\ 
If we now compare MC-srOEP with the MCOEP
approach of Weimer {\it et al.}~\cite{JCP08_Gorling_mcoep}, they differ in many respects. First, the MCOEP
wavefunction is a linear combination of determinants 
constructed from KS-OEP-type orbitals. As a result, the optimization of
the OEP requires the computation of the  
linear response of KS determinants related to changes in the potential,
for which simple analytical expressions can be derived. In this
respect, MC-srOEP is more complicated to implement, as the linear response of an
MC long-range interacting wavefunction is required.\\ 
Second, 
an important difference between MC-srOEP and MCOEP methods relates to the active
space. In the exact formulation of the MCOEP method, the density constructed from the MCOEP wave
function for a fixed active space is equal to the exact ground-state density of the
physical system and the exact ground-state energy is recovered.
Therefore, the 
complementary density-functional correlation term, which
describes dynamical correlation, depends on the active space. Developing
approximate density functionals for this scheme is a difficult
task as correlation effects may be double counted.\\ 
On the other hand, the ``exact" MC-srOEP wavefunction gives the exact energy and density 
only in the FCI limit in an infinite
basis set. Nevertheless, with relatively small $\mu$ values and the same active space as in the MCOEP expansion, the ``exact" MC-srOEP
density and energy should be
close to the exact density and energy of the physical system, given that
short-range effects are described by the exact MD
short-range correlation functional and long-range effects are 
treated exactly in the given active space.\cite{PaolasrXmd} As already pointed out, the
advantage of such a scheme is that the complementary MD
short-range correlation density-functional does not depend on the active space,
making it easier to model. In this work, the MD srLDA functional of Paziani {\it et al.}~\cite{srmdCref} is used.   

The last approximation discussed here concerns the srOEP parameterization. 
Following Wu and Yang,\cite{Yang_prl_oep} we introduce an expansion of the potential
\begin{align}\label{vgtexp}
v(\mathbf{r})= v_{\text{ne}}(\mathbf{r}) + v^{\rm sr,\mu}_{\text{ref}}(\mathbf{r}) + \sum_t b_t g_t(\mathbf{r}),
\end{align}  
where the short-range analogue of the Fermi--Amaldi potential, calculated
for a fixed $N$-electron density $n_0$, is employed as the reference potential:
\begin{align}\label{vgaussian}
 v_{\text{ref}}^{{\rm sr},\mu}({\bf r})&= \frac{N-1}{N}\int {\rm d}{\bf r'}\; n_0({\bf r'}){w}^{\rm sr,\mu}_{\rm ee}(|{\bf r}-{\bf r'}|).
\end{align}
We use the same basis set $\{g_t\}$ for the expansion of the potential 
and the molecular orbitals.
This parameterization allows for the use of analytic derivatives in quasi-Newton
approaches to perform the optimization of Eqs.~(\ref{GsenevmuHF}) and
(\ref{GsenevmuMC}), and thus
determine the potential expansion coefficients $\{b_t\}$. As a first
step, we here present the derivation of the HF-srOEP
gradient. The implementation of the analytical MC-srOEP gradient is in
progress and will be presented in a separate paper. The MC-srOEP results
presented in Sec.~\ref{subsubsec:mc-sroep-h2} for H$_2$ were obtained numerically, by
finite differences.

\subsection{Analytical HF-srOEP energy gradient}\label{subsecHFOEP}

The computation of the HF-srOEP energy in Eq.~(\ref{GsenevmuHF}) can be performed
with quasi-Newton approaches, using the coefficients $\{ b_t\}$ in the
potential expansion of Eq.~(\ref{vgtexp}) as variational parameters. 

Let $v_0$ denote the trial
potential defined by the initial set of
coefficients $\{ b^{(0)}_t\}$. The associated determinant
$\Phi^{\mu}[v_0]$ in Eq.~(\ref{phimuvfromminhf}) is denoted
$\Phi^{\mu}_0$ in the following. Variations in the potential
coefficients 
\begin{align}
v_0(\mathbf{r})\rightarrow v_0(\mathbf{r})+\sum_t \epsilon_t g_t(\mathbf{r})
\end{align}
can be interpreted as static perturbations, where the property operators
are the Gaussians $g_t$ with perturbation strengths
$\epsilon_t$. We denote by $i,j$ and $a,b$ the occupied and unoccupied real-valued
orbitals in $\Phi^\mu_0$, respectively. We use a second-quantized exponential
parameterization~\cite{hf_pinkbook} for the
determinant $\Phi$ in Eq.~(\ref{phimuvfromminhf}),  
\begin{align}\label{HFwf}
\vert\Phi({\boldsymbol \kappa})\rangle=e^{-\hat{\kappa}}|\Phi^{\mu}_0\rangle,
\end{align}
where
\begin{align}\label{kappaoperator}
\hat{\kappa}&=\sum_{a,i}\kappa_{ai}\left(\hat{E}_{ai}-\hat{E}_{ia}\right), \nonumber \\
\hat{E}_{ai}&=\hat{a}^{\dagger}_{a,\alpha}\hat{a}^{\phantom{\dagger}}_{i,\alpha} +\hat{a}^{\dagger}_{a,\beta}\hat{a}^{\phantom{\dagger}}_{i,\beta}.
\end{align}
The HF-srOEP energy gradient can then be expressed in terms of the orbital
rotation vector 
\begin{align}\label{orbrotvec}
{\boldsymbol \kappa}= 
\begin{bmatrix}
\vdots\\
\kappa_{ai}\\
\vdots\\
\end{bmatrix}
\end{align}
as follows:
\begin{align}\label{gradientE}
\left . \frac{{\rm d}E}{{\rm d}\epsilon_t}\right |_{0}= \left .\frac{\partial E}{\partial {\boldsymbol \kappa}}\right |_{0}^{\rm T} \left .\frac{\partial{\boldsymbol \kappa}}{\partial \epsilon_t}\right |_{0},  
\end{align}
with, according to Eq.~(\ref{GsenevmuHF}),
\begin{align}\label{hf-sroepener}
E(\boldsymbol \kappa) =\langle \Phi({\boldsymbol \kappa}) \vert \hat{T}+\hat{W}_{\rm ee}+\hat{V}_{\rm ne}\vert \Phi({\boldsymbol \kappa}) \rangle +E^{\rm sr,\mu}_{\rm c, md}[n({\boldsymbol \kappa})],
\end{align}
and 
\begin{align}
n({\boldsymbol\kappa},{\mathbf r})=\langle \Phi({\boldsymbol\kappa}) \vert\hat{n}({\mathbf r})\vert\Phi({\boldsymbol\kappa})\rangle.
\end{align}
Note that the Hellmann--Feynman theorem cannot be applied 
in this context since the HF-srOEP energy depends implicitly on the potential.
By analogy with regular HF theory,\cite{hf_pinkbook}
the energy gradient components can be written as 
\begin{align}\label{frenergradient}
\left .\frac{\partial E}{\partial {\kappa_{ai}}}\right |_{0} =-4 \left(f_{ai} +\langle a\vert \hat{v}^{\rm sr, \mu}_{\rm c, md}[n_{\Phi^{\mu}_0}]\vert i\rangle\right),  
\end{align}
where $f_{ai}$ is the conventional Fock-matrix element computed with
HF-srOEP orbitals, while the MD short-range correlation
density-functional potential 
${v}^{\rm sr, \mu}_{\rm c,
md}[n_{\Phi^{\mu}_0}](\mathbf{r})={\delta E^{\rm sr,\mu}_{\rm c,
md}}/{\delta n(\mathbf{r})}[n_{\Phi^{\mu}_0}]$
is calculated for the HF-srOEP density. 
As shown in the Appendix, the linear response vector is obtained as
follows   
\begin{align}\label{linearrspeqfinal}
\mathcal{E}^{[2]\mu} \left .\frac{\partial{\boldsymbol\kappa}}{\partial \epsilon_t}\right |_{0}=-g^{[1]}_t,
\end{align}
where the gradient property vector is equal to 
\begin{align}\label{gtpropvec}
g^{[1]}_t= -4
\begin{bmatrix}
\vdots\\
\langle a |\hat{g}_t| i\rangle\\
\vdots\\
\end{bmatrix}.
\end{align}
The long-range analog of the HF Hessian, $\mathcal{E}^{[2]\mu}$, is 
in the canonical HF-srOEP orbital basis equal to~\cite{hf_pinkbook} 
\begin{align}\label{lrHFhessian}
\mathcal{E}_{ai,bj}^{[2]\mu}=4 &\left(  \delta_{ab} \delta_{ij} \left(\varepsilon^\mu_a-\varepsilon^\mu_i \right) +4 \langle ab\vert ij\rangle^{\rm lr,\mu}  \right. \nonumber \\
&  \left. -\langle ai\vert bj\rangle^{\rm lr,\mu} - \langle ai\vert jb\rangle^{\rm lr,\mu} \right),
\end{align}
where $\varepsilon^\mu_a$ and $\varepsilon^\mu_i$ are the unoccupied and
occupied HF-srOEP orbital energies, respectively. We conclude from
Eqs.~(\ref{gradientE}) and (\ref{linearrspeqfinal}) that the HF-srOEP
energy gradient can be written as
\begin{align}\label{gradientE2}
\left . \frac{{\rm d}E}{{\rm d}\epsilon_t}\right |_{0}=-\left .\frac{\partial E}{\partial {\boldsymbol \kappa}}\right |_{0}^{\rm T}\left[\mathcal{E}^{[2]\mu}\right]^{-1}g^{[1]}_t,  
\end{align}
or, equivalently,
\begin{align}\label{gradientE_invhessian2}
\left . \frac{{\rm d}E}{{\rm d}\epsilon_t}\right |_{0}=\overline{\boldsymbol \kappa}^{\rm T}\,g^{[1]}_t,  
\end{align}
where $\overline{\boldsymbol \kappa}$ fulfills the linear response equation
\begin{align}\label{linearrspkappabar}
\mathcal{E}^{[2]\mu}\,\overline{\boldsymbol \kappa}=-\left .\frac{\partial E}{\partial {\boldsymbol \kappa}}\right|_{0}.
\end{align}
Note that all components of the HF-srOEP energy gradient 
are thus computed from 
one single linear response
vector $\overline{\boldsymbol \kappa}$. The latter can be obtained 
straightforwardly from a standard second-order HF wavefunction
optimizer~\cite{hf_pinkbook} when (i) using long-range integrals and substituting the
trial srOEP for the nuclear potential in the Hessian and (ii) adding the
MD short-range density-functional potential calculated for the
HF-srOEP density to the Fock operator in the energy gradient.\\ 
We finally mention that, at $\mu=0$, the long-range integrals are
zero and the MD short-range correlation density-functional potential reduces to the
conventional correlation density-functional potential $v_c[n](\mathbf{r})$. As a result,  
the HF-srOEP determinant becomes the standard KS-OEP determinant
$\Phi^{\rm KS}$, the orbital energies reduce to
conventional KS-OEP energies
$\varepsilon_a$ and $\varepsilon_i$, and the   
Hessian becomes diagonal:
\begin{align}\label{lrHFhessianmuzero}
\mathcal{E}_{ai,bj}^{[2]0}&=4\delta_{ab}\delta_{ij}\left(\varepsilon_a-\varepsilon_i\right).
\end{align}
We thus obtain from Eqs.~(\ref{frenergradient}), (\ref{gtpropvec}) and
(\ref{gradientE2}) the following analytical expression for the energy gradient 
\begin{align}\label{gradientErspKS}
\left . \frac{{\rm d}E}{{\rm d}\epsilon_t}\right|_{0} \underset{\mu=0}{\longrightarrow} \hspace{2mm} 4\sum_{a,i} & \frac{\langle a\vert \hat{g}_t\vert i\rangle}{\varepsilon_i-\varepsilon_a} \nonumber\\
&\times \left(f_{ai} +\langle a\vert \hat{v}_{\rm c}[n_{\Phi^{\rm KS}}]\vert i\rangle\right).
\end{align}
When the correlation potential is neglected, the KS-EXX energy gradient expression of Yang and Wu\cite{Yang_prl_oep} is recovered. 

\subsection{Summary}\label{subsec:summary}

In conventional MC-srDFT the KS decomposition of the complementary short-range
exchange--correlation density-functional energy is used. Within the
local density approximation,\cite{toulda} the scheme will be referred to as MC-srLDA. The
alternative separation of exchange and correlation energies that we
investigate in this work relies on the multi-determinantal (MD) long-range interacting
wavefunction rather than the KS determinant. As a result, long-range and
short-range interactions can be recombined in the energy expression.
The latter is thus rewritten as the sum of the expectation value for the
regular Hamiltonian and a complementary short-range
density-functional correlation energy that is referred to as MD.
The long-range MC wavefunction to be inserted into this energy
expression cannot be obtained straightfowardly by minimization over the
wavefunction parameters otherwise double counting occurs. 

Various approximations utilising the alternative MD decomposition are considered in this work. The simplest consists of using
the MC-srLDA wavefunction. This approximation is referred to as RSMCHf.
We may also consider, for analysis purposes, the single-determinant
version of RSMCHf, that we refer to as RSHf and which consists of
computing the energy with the HF-srLDA determinant rather than the
MC-srLDA wavefunction. A more
sophisticated procedure uses short-range OEPs. These can be optimized
either at the HF or MC levels, leading to the HF-srOEP and MC-srOEP
models. The analytical energy gradient has been
derived and implemented for HF-srOEP. The implementation of the
analytical MC-srOEP
gradient is currently in progress. For analysis purposes, the
long-range MC wavefunction can still be computed without
reoptimization of the srOEP. This scheme,
where a 
frozen effective
potential (FEP) is employed, will be referred to as MC-srFEP. The
HF-srOEP potential has been used as the srFEP in the following.

Working equations associated with all these schemes are given in Table~\ref{summary_methods}.  

\section{Computational details}\label{comp_details_sec}

The various range-separated DFT schemes listed in
Sec.~\ref{subsec:summary} have been
implemented in a development version of the
\textsc{\texttt{DALTON}}2011 program.\cite{daltonpack}~The MD 
srLDA correlation functional of Paziani~{\it et al.}~\cite{srmdCref} has
been used. MC-srLDA wavefunctions and energies have been computed with the srLDA exchange--correlation
functional of Toulouse~{\it et al.}~\cite{toulda}
For the HF-srOEP and MC-srOEP approaches the minimizations of
Eq.~(\ref{GsenevmuHF}) and Eq.~(\ref{GsenevmuMC}), respectively, were
performed using the Broyden--Fletcher--Goldfarb--Shanno 
(BFGS) quasi-Newton algorithm\cite{Yang_prl_oep}. The initial Hessian was taken to be the approximate Hessian expression of Ref.~\onlinecite{Yang_Jthcompchem_oep}. Convergence to a gradient norm below $10^{-5}$ or a maximum absolute change in potential coefficients of less than $10^{-12}$ is typically achieved in less than 20 iterations with this choice. 
For the HF-srOEP approach the energy gradient required at each iteration is
computed according to Eqs.~(\ref{gradientE_invhessian2}) and
(\ref{linearrspkappabar}).

For the MC-srOEP approach, analytical gradients are not yet implemented (though we have derived their form), however, for analysis purposes in the present work calculations are carried out by determining the required gradient by finite difference. Since this substantially increases the number of energy evaluations to be performed we have considered the fully optimized MC-srOEP approach only for the H$_2$ molecule.
Potential energy curves, equilibrium bond lengths and dissociation
energies have been calculated for H$_2$, Li$_{2}$, N$_2$ and H$_{2}$O. 
All calculations were performed with uncontracted cc-pVTZ basis
sets~\cite{basissets1, basissets2} for both the orbital and potential
expansions. Un-contraction of the basis sets and the use of the same
sets for each expansion ensures smooth physically reasonable srOEP
potentials are obtained. The active orbital spaces used in the
multi-configuration calculations are
1$\sigma_g$1$\sigma_u$ for H$_2$ and
2$\sigma_g$2$\sigma_u$1$\pi_u$3$\sigma_g$1$\pi_g$3$\sigma_u$ for N$_2$
and Li$_2$. For H$_2$O, the active orbital space is denoted 3.1.2.0, which
signifies the number of orbitals in the a$_1$.b$_1$.b$_2$.a$_2$
symmetries, respectively. The $C_2$ symmetry axis is along the $z$ axis,
and the $\sigma_v$ and $\sigma'_v$ mirror planes are $\sigma_v(xz)$ and
$\sigma_v(yz)$, respectively.  
\section{Results and discussion}\label{sec:results}

\subsection{Choice of the $\mu$ parameter}\label{subsec:choice_mu_parameter}

In the context of range-separated hybrid functionals, where
range separation is used for the exchange energy only, the $\mu$
parameter is usually optimized semi-empirically for thermochemistry and
other desired properties, leading thus to an average system-independent value
in the range $0.4$--0.5.
\cite{hiraomu,hiraonewmu,scuseriacalib1,scuseriacalib2,nancycalib,headgordonmu,dft-Rohrdanz-JCP2009-130-054112}
 Quite recently, Baer
{\it et al.}~\cite{PRL10_Baer_tuning_mu_Koopmans} proposed  
to choose $\mu$ such that Koopmans' theorem for
both neutral and anion is obeyed, as closely as possible. This
procedure, which relies on first principles,
enables one to tune the $\mu$ parameter for a given system. Let us stress
that, in the exact theory, any $\mu$ value would provide the same
(exact) ground-state energy. The problem of choosing $\mu$ occurs in
practice because approximate short-range exchange functionals are used.  
In the context of 
multi-determinant range-separated DFT, where range separation is used
for both exchange and correlation energies, the optimal choice of $\mu$ is even
more problematic as, in practice, both approximate wavefunctions and density functionals are
employed. \'{A}ngy\'{a}n and coworkers~\cite{srdftnancy,JCP07_Ian_mp2-srdft_calib_Rg1-Rg2,JCP10_Wuming_rpa-srdft_weak_int,rpa-srdft_toulouse_2011,PRA10_Julien_rpa-srDFT} 
use for example $\mu=0.5$ in their range-separated {\it second-order M{\o}ller-Plesset} (MP2) or {\it random phase approximation} (RPA)
calculations on weakly interacting
systems.
This value has been calibrated in a completely different context, that
is the exchange-only
range-separated hybrid one, for reproducing atomization energies
of small molecules.~\cite{nancycalib} In the particular case of the
homonuclear rare-gas dimers, Goll~{\it et al.}
~\cite{
ccsrdft} alternatively proposed to choose for 
$\mu$ the inverse of the van der Waals radius
in their range-separated {\it coupled cluster} (CC) calculations, so that
intra-atomic correlations could be essentially treated in DFT.

In the context of MC-srDFT, 
Fromager \textit{et
al.}~\cite{dft-Fromager-JCP2007a} investigated the possibility of choosing
$\mu$ in such a way that static and dynamical correlations could be assigned to
the long-range MCSCF and short-range density-functional correlation energies, respectively. As static correlation is usually not a purely
long-range effect, even in the simple case of the dissociated H$_2$
molecule,~\cite{AMTOptseplrsr} the authors focused on the dynamical correlation, suggesting 
that $\mu$ should be chosen small enough that the Coulomb hole is essentially treated in DFT.
On the other hand it must also be chosen large enough that, in cases
where static correlation becomes significant, the wave function can
become sufficiently multi-configurational. The authors
proposed from these considerations the following prescription: the
largest value of $\mu$ for which the MC-srDFT wave
function is well approximated by a single determinant, in  
systems where static
correlation is not significant, should be considered as optimal. Let us stress that such a prescription
does not guarantee that MC-srDFT will perform well when applied to
systems with static correlation. It only ensures that the Coulomb hole is
described within DFT and that the long-range part of the static
correlation is assigned to MCSCF. 

Obviously, with this choice, the
complementary short-range correlation density-functional is
expected to model the short-range part of the static correlation. As
discussed further in Sec.~\ref{sec:results}, this
can be problematic when stretching a bond for example. Numerical values for $\mu$ were obtained when analyzing
long-range correlation effects as $\mu$ varies. Two strategies were proposed. The
first one is based on the energy and consists of examining the total energy difference $\Delta E_{\rm c}^\mu$ between the MC and HF approximations for a
given range-separated scheme. The best $\mu$ value is then determined by examining $\Delta E_{\rm c}^\mu$ for systems dominated by dynamical correlation and choosing the
$\mu$ value at which this quantity falls below a threshold of -1 $mE_h$. The second strategy examines, for the same systems, the natural orbital
occupancies within the MC approximation and choosing the $\mu$ value at
which these deviate from 2 (with a threshold of $10^{-4}$). Calculations
on a small test set of systems containing light elements all yielded the
optimal $\mu=0.4$ value.

\begin{figure}
{\includegraphics[width=0.7\textwidth
]{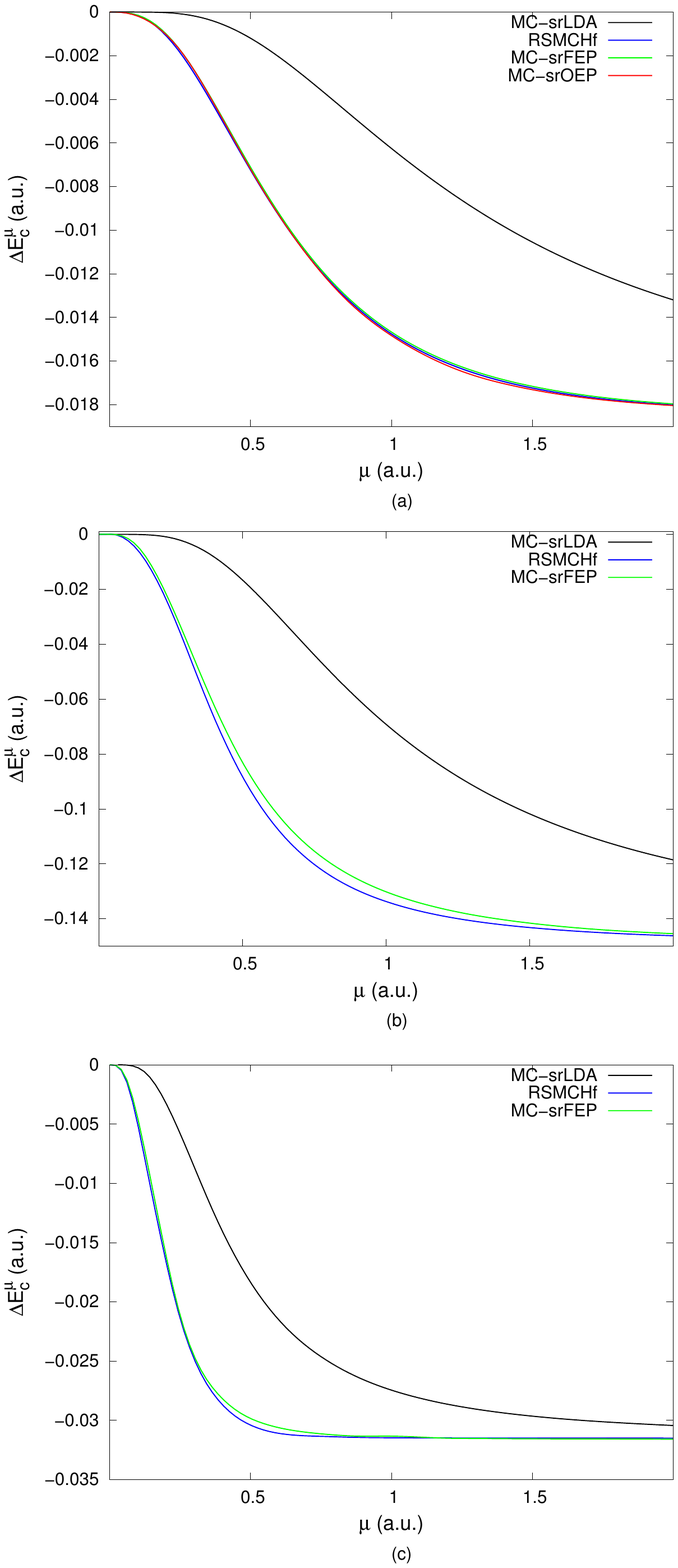}}
\caption{\label{deltaEmu_curves_all} The quantity $\Delta E_{\rm c}^\mu$ for (from top to bottom) H$_2$, N$_2$ and Li$_2$ for each method as a
function of the parameter $\mu$. See text for further details.}
\end{figure}

We now investigate whether this value is still optimal when a
different separation of exchange and correlation energies is
employed. We use for the discussion the H$_2$ molecule in its
equilibrium geometry ($R_e=
0.741$
\AA~\cite{lie1,lie2}
) as an example of system that is completely dominated by dynamical
correlation. 
N$_2$ and Li$_2$ in
their equilibrium geometries ($R_e=$1.097 \AA~\cite{lie1} and 2.673
\AA~\cite{liexpdist, liexpdist2}, respectively) will then be considered.
These systems are interesting as they exhibit, Li$_2$ in particular, a
multi-configurational character already at
equilibrium. As discussed in the following, it is of course not as pronounced as
in the dissociation limit but it is not negligible. Following
Ref.~\onlinecite{dft-Fromager-JCP2007a}, we have computed the energy difference $\Delta
E_{\rm c}^\mu$ at the RSMCHf level. Results are
shown in Fig.~\ref{deltaEmu_curves_all}. At the RSMCHf level of theory, $\Delta
E_{\rm c}^\mu$ deviates 
from zero (to within 10$^{-3}$~a.u.) for much smaller $\mu$ values than at the MC-srLDA level. 
For H$_2$, for example, the RSMCHf and MC-srLDA $\mu$ values are 0.25 and 0.5, respectively. 

It is tempting to conclude that the prescription of Fromager
\textit{et al.}~\cite{dft-Fromager-JCP2007a} leads to different optimal
$\mu$ values when considering RSMCHf energies. The situation is more
subtle, however. The $\Delta E_{\rm c}^\mu$ energy
difference calculated within RSMCHf contains different
correlation effects to the one obtained within the MC-srLDA scheme.
In the latter case, $\Delta E_{\rm c}^\mu$ is essentially the purely long-range correlation energy
while, within RSMCHf, it also contains the coupling between short- and
long-range correlations that arises from the expectation value
of the short-range interaction over the long-range correlated MC-srLDA
wavefunction (see Eq.~(\ref{energymin3})). This appears clearly in
range-separated {\it second-order density-functional perturbation
theory} (DFPT2) where a
long-range MP2 description is used rather than a MCSCF
one.\cite{Doublehybmp2_yann}
For small $\mu$ values,
the difference between MC-srLDA and RSMCHf correlation energies can be rationalized by considering  
the Taylor expansions of the long- and short-range interactions,
\begin{align}\label{taylorexplr}
w^{\rm lr,\mu}_{\rm ee}(r_{12})
&=\frac{2}{\sqrt{\pi}}\left(\mu-\frac{1}{3}\mu^3r_{12}^2+\mathcal{O}(\mu^5)\right), \nonumber\\
w^{\rm sr,\mu}_{\rm ee}(r_{12})&=\frac{1}{r_{12}}+\mathcal{O}(\mu).
\end{align}
As the deviation of the exact long-range interacting wavefunction from
the KS determinant 
varies as
$\mu^3$ (see Appendix B4~of Ref.~\onlinecite{erferfgaufunc}), the long-range correlation energy
varies as $\mu^6$ and the coupling between long- and short-range
correlations as $\mu^3$. As a result, the latter is expected to
deviate more rapidly from zero as $\mu$ increases. 
This is the reason why a
threshold of $-1 mE_h$ on $\Delta E_{\rm c}^\mu$ will provide a smaller $\mu$ value for RSMCHf than for MC-srLDA.  
As shown in Fig.~\ref{deltaEmu_curves_all}, the same conclusion can be drawn for 
MC-srFEP and MC-srOEP. This was expected as all three models use
the same energy expression. They only differ by the long-range MC
wavefunction that is inserted into this expression for the computation
of the energy. 
If we want to follow the prescription of Fromager {\it et al.},
~\cite{dft-Fromager-JCP2007a} which relies on the
analysis of purely long-range correlation effects, one should 
extract from $\Delta E_{\rm c}^\mu$ the purely long-range correlation
energy or increase the threshold. As this analysis is
performed on static-correlation free electronic systems, range-separated
DFPT2~\cite{Doublehybmp2_yann} is
expected to be a good approximation to RSMCHf, especially for small
$\mu$ values. In this context, the coupling between long- and
short-range correlations can be separated from the purely long-range
correlation energy and each term can be computed when varying $\mu$.
Results obtained for rare gas atoms are presented in
Ref.~\onlinecite{Doublehybmp2_yann}. When $\mu=0.4$, the coupling term
equals -4 and -20 $mE_h$
 in He and Ne, respectively, while the long-range correlation energy
is above -1 $mE_h$ in both systems. This suggests that the coupling term
is more system-dependent than the purely long-range correlation energy.
It was 
expected as the former is expressed in terms of both long- and
short-range integrals. 
In order to reduce the system-dependency of the $\mu$ parameter, one may
want to examine purely long-range correlation effects only, leading thus to the optimal $\mu=0.4$
value for the rare gas atoms~\cite{Doublehybmp2_yann}.       

\begin{figure}
\includegraphics[width=0.7\textwidth
]{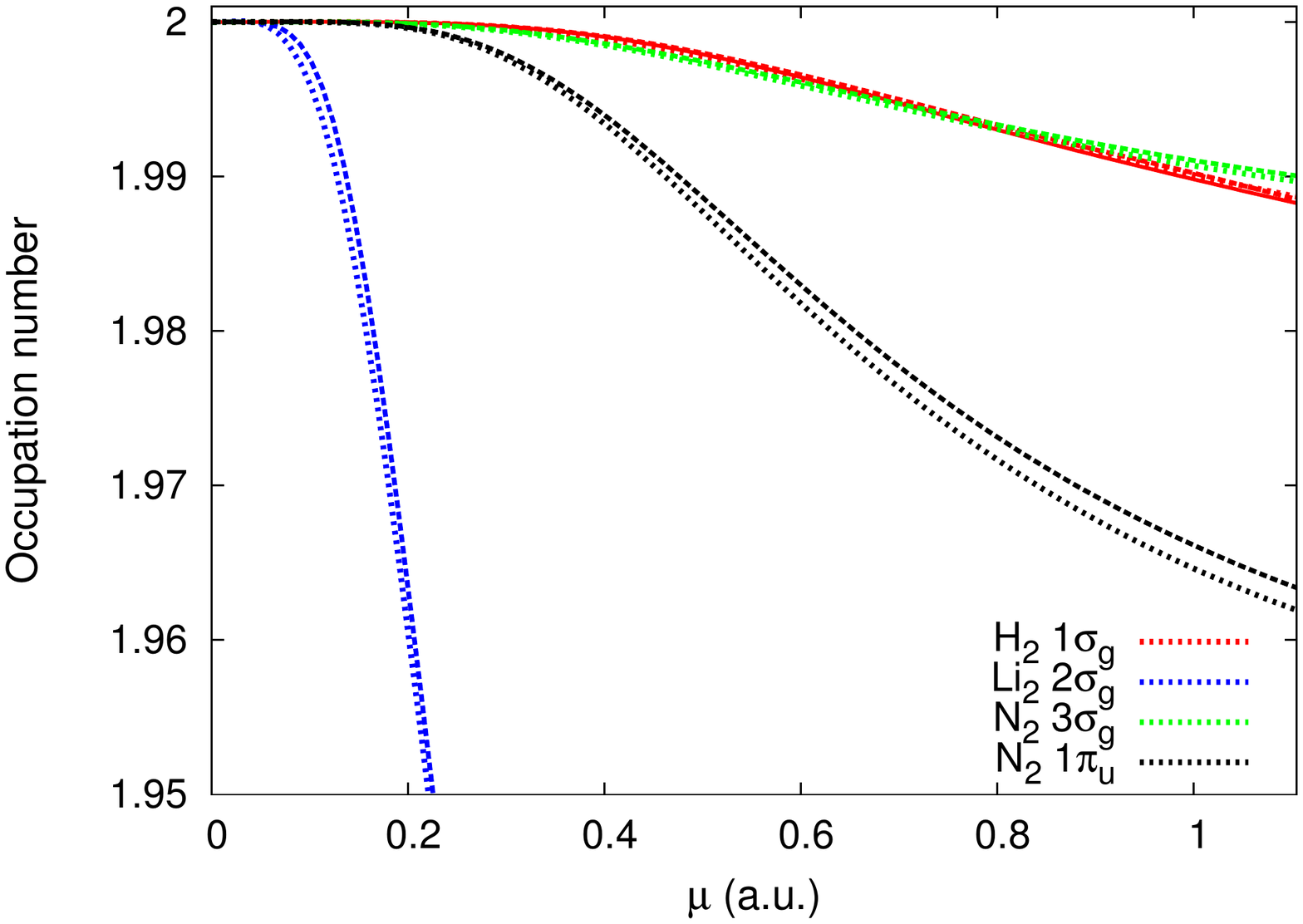}
\caption{\label{Occnwithmu_curves} Occupation numbers of MC-srLDA (dotted), MC-srFEP (dashed) and MC-srOEP (solid)
active natural orbitals as a function of $\mu$ for H$_2$, N$_2$ and Li$_2$ at their experimental equilibrium geometries~\cite{lie1,lie2,liexpdist,liexpdist2}.}
\end{figure}

Returning to H$_2$ and the RSMCHf model, we can still utilize, as an alternative, the second strategy of
Ref.~\onlinecite{dft-Fromager-JCP2007a}
that relies on the analysis of the  
natural orbital occupancies as $\mu$ increases from zero. Since RSMCHf and MC-srLDA
wavefunctions are identical by definition, we can simply refer to
Ref.~\onlinecite{dft-Fromager-JCP2007a}
and conclude that $\mu=0.4$ is also optimal in this context.
This is illustrated by the MC-srLDA occupancies of H$_2$ in
Fig.~\ref{Occnwithmu_curves}. Interestingly, similar conclusions can be
drawn for H$_2$ at both MC-srFEP and MC-srOEP levels. This should
clearly be investigated on more static-correlation-free systems, once
the analytical MC-srOEP energy gradient is implemented. This is left for future work. 
Note that $\mu=0.4$ is large enough to assign static correlation in
Li$_2$ and N$_2$, or at
least a part of it, to the long-range MCSCF as suggested by their natural orbital occupancies in Fig.~\ref{Occnwithmu_curves}.

In summary, our
preliminary calculations suggest that 
it is relevant to use 
$\mu=0.4$ in conjunction with 
the alternative separation of exchange and correlation energies of 
Toulouse {\it et
al.}\cite{TousrXmd}  

\subsection{Potential curves of H$_2$, N$_2$, Li$_2$ and H$_2$O}

\subsubsection{RSMCHf equilibrium distances and binding energies}\label{subsubsec:rsmch_results}

\begin{figure*}
{\label{Fig3}\includegraphics[width=0.85\textwidth
]{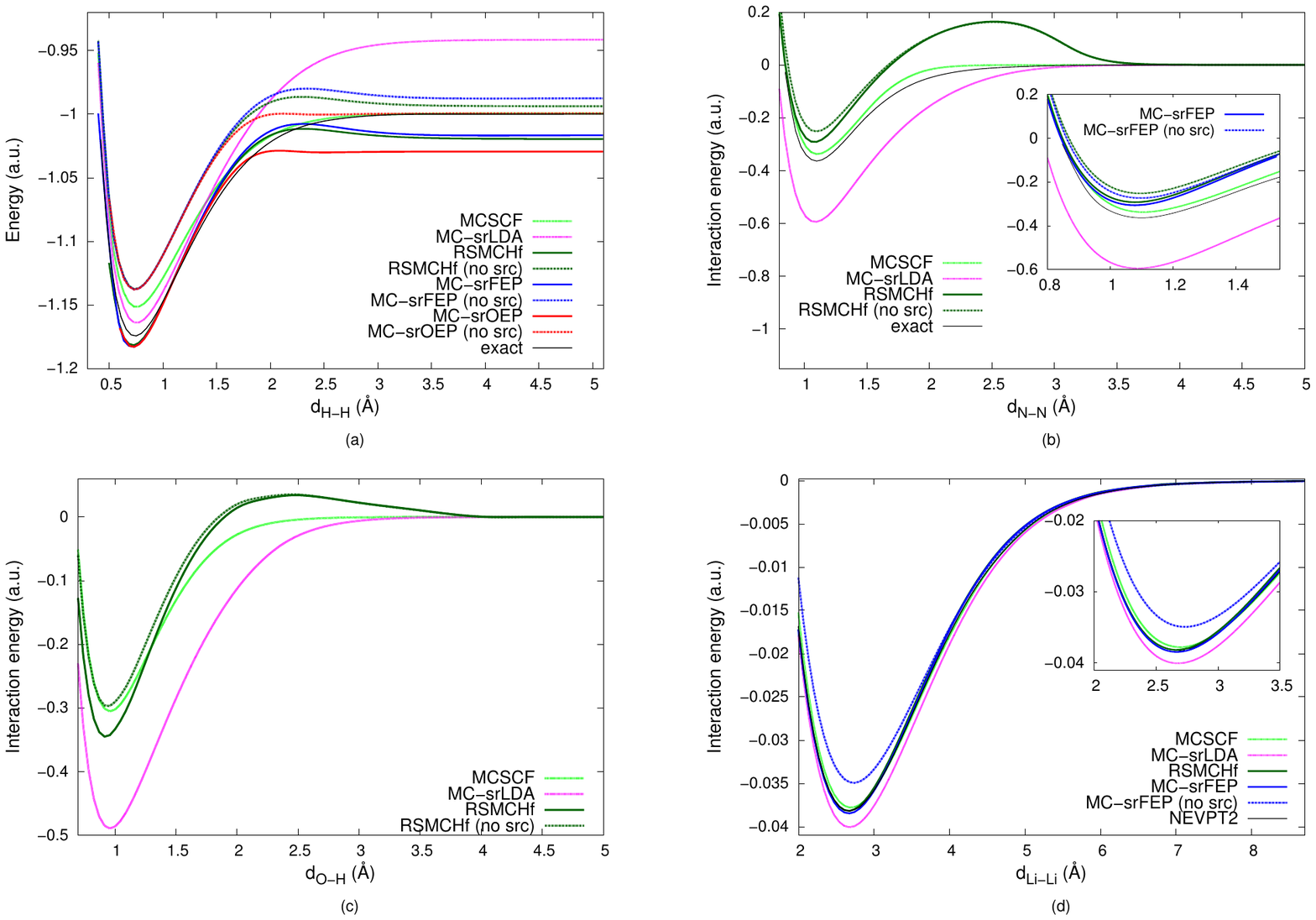}}
\caption{\label{PECandIE_curves} Potential energy curves (a.~u.) of H$_2$ (upper left panel) and interaction (binding) energies of N$_2$ (upper right panel), H$_2$O (lower left panel) and Li$_2$ (lower right panel) obtained by means of the new multi-configuration range-separated schemes. The coupling parameter is $\mu$=0.4. The exact interaction energy and PEC curves are taken from Ref.~\onlinecite{lie2} for H$_2$ and Ref.~\onlinecite{lie1} for N$_2$.}
\end{figure*}

In this section, we compare the performance of the RSMCHf 
and MC-srLDA schemes for the description of the potential energy curves (PECs) 
of H$_2$, N$_2$, Li$_2$ and H$_{2}$O. In the latter case,
the symmetric dissociation at the experimental
equilibrium angle H--O--H of 104.5$^\circ$~\cite{h2oexper} was investigated.
PECs as well as equilibrium bond distances and dissociation energies are
given in Fig.~\ref{PECandIE_curves} and Tables~\ref{ReDeRe} and~\ref{ReDeRe_second_half},
respectively. According to  
Sec.~\ref{subsec:choice_mu_parameter} the $\mu$ parameter
has been set to 0.4.

For the H$_2$ molecule, close to the equilibrium geometry, MC-srLDA
has a total energy that is too positive, though it recovers more
short-range dynamical correlation than standard MCSCF. Although the
qualitative shape of the curve is much better than restricted
Hartree--Fock (not shown), the energy in the dissociation regime is much
too positive. Overall this leads to a dissociation energy which is much
too large, as shown in Table~\ref{ReDeRe}. The RSMCHf approach gives
energies close to equilibrium that are slightly too negative, whilst
those at dissociation are significantly too negative. The result is a
slight underestimation of the dissociation energy. 
The deviations from the exact curve in the dissociation regime come from the complementary MD srLDA 
correlation functional, as illustrated by the RSMCHf (no src)
PEC in Fig.~\ref{PECandIE_curves}, obtained when subtracting the former from the RSMCHf energy.
As expected, the RSMCHf (no src) energy, which is equal to the 
expectation value of the regular Hamiltonian over the MC-srLDA wavefunction,
is greater than the pure MCSCF energy for all bond distances. 
Returning to the equilibrium distance, the too negative MD srLDA correlation energy for H$_2$ 
at $\mu=0.4$ was already observed by Gori-Giorgi and Savin~(see the uppermost panel in Fig.~7 of
Ref.~\onlinecite{PaolasrXmd}),
who obtained an error (about 0.01$E_\text h$~in absolute value)
close to that at the RSMCHf level (0.007$E_\text h$)\ when
comparing with the ``exact" energy of Ref.~\onlinecite{lie2}.
In spite of these errors, the RSMCHf model clearly
improves upon the dissociation energy obtained at the MC-srLDA level,
reducing the absolute error from 27 to 7\% for H$_2$, with similar
conclusions for the other systems.

\begin{figure*}
{\includegraphics[width=0.7\textwidth
]{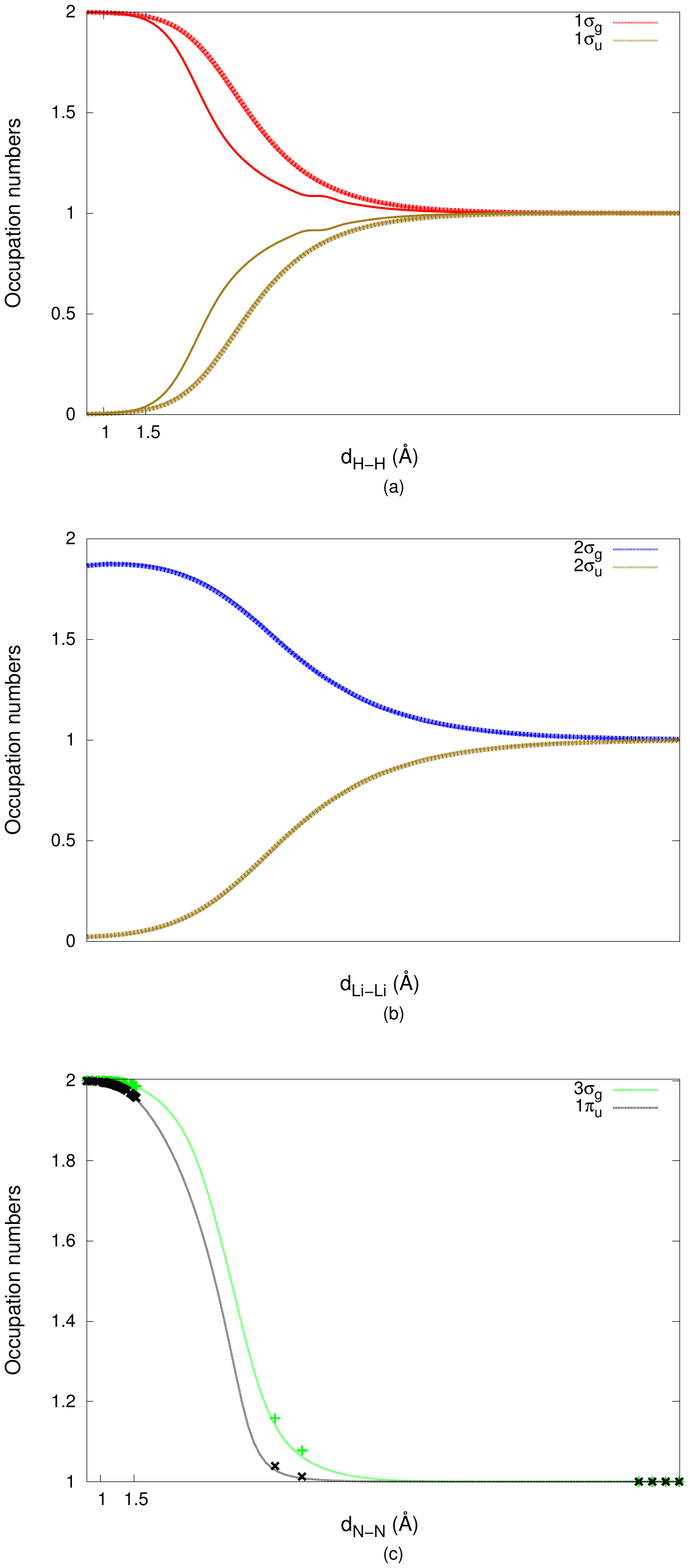}}\\
\caption{\label{Occnumberswithdist} Occupation numbers of MC-srOEP (solid), MC-srFEP (dotted) and MC-srLDA (dashed) active natural orbitals as a function of bond distance (\AA) (from top to bottom) H$_2$, Li$_2$ and N$_2$. The parameter $\mu$ is set to 0.4.}
\end{figure*}

The calculated equilibrium bond distances compare relatively well with experiment at both MC-srLDA and RSMCHf
levels (see Table~\ref{ReDeRe}). While the MC-srLDA model slightly
overestimates the bond distance of H$_2$ by 0.015 \AA, 
the RSMCHf model underestimates it by 0.02\,\AA.
This slight over-binding is induced by the complementary
MD srLDA correlation functional, as suggested by the RSMCHf (no src) equilibrium distance of 0.744\,\AA, 
which is almost equal to the experimental value. 
Note that MC-srLDA and HF-srLDA bond distances are almost identical
since, for $\mu=0.4$, the MC-srLDA wave
function is well approximated by a single determinant (see
Sec.~\ref{subsec:choice_mu_parameter} and Fig.~\ref{Occnumberswithdist}).

On the other hand, RSHf and RSMCHf equilibrium
distances are quite similar but not as close as HF-srLDA and MC-srLDA
distances (see Table~\ref{ReDeRe}). The (slight) difference may be
caused by the short/long-range correlation coupling, which contributes
significantly to the RSMCHf correlation energy, where it is treated
explicitly within the MCSCF, unlike for the MC-srLDA energy, where it is 
described within DFT (see Sec.~\ref{subsec:choice_mu_parameter}). 
Similar conclusions can be drawn for the other systems when
comparing MC-srLDA with RSMCHf
equilibrium distances, as shown in Tables~\ref{ReDeRe} and
\ref{ReDeRe_second_half}. 

\subsubsection{The treatment of static correlation within RSMCHf}

Regarding the H$_2$ dissociation limit, the large error in the MC-srLDA
energy
was interpreted in Ref.~\onlinecite{dft-Fromager-JCP2007a} as a
self-interaction error of the complementary spin-unpolarised srLDA exchange--correlation functional. 
We emphasize that the complementary short-range correlation energy, whose exact expression is given in Eq.~(\ref{ACsrcfun}), 
is not supposed to be zero at large internuclear 
distances---instead, it should compensate the short-range Hartree and exchange energy
contributions. This was shown by Teale {\it et al.,}\cite{AMTOptseplrsr}   
who computed accurately the correlation integrand in Eq.~(\ref{srcintegrand}) for various bond distances. 
According to Eq.~(\ref{ACsrcfun}), 
 the accurate short-range correlation energy is obtained for $\mu=0.4$
 when integrating the correlation integrand in Fig.~6 (f) of
 Ref.~\onlinecite{AMTOptseplrsr} from
 $\mu/(1+\mu)=0.286$ to 1. This quantity is obviously not equal to zero. The sum of
short-range Hartree, exchange and correlation energies should, on
 the other hand, vanish in the dissociation limit. The corresponding
 integrand in Eq.~(\ref{srHxcintegrand}) is plotted in Fig.~6 (d) of
 Ref.~\onlinecite{AMTOptseplrsr}. Pure short-range exchange--correlation
 density functionals like srLDA are simply unable to compensate the short-range
 Hartree term.~\cite{dft-Fromager-JCP2007a}
 
The error in the total energy 
at large distances is significantly reduced when using the RSMCHf model,
based on a different decomposition of
 the short-range exchange--correlation energy. The complementary
 short-range MD density-functional
 correlation energy is defined, in RSMCHf, with respect to the long-range
 interacting wavefunction rather than the KS non-interacting one. Its
 exact expression is given in Eq.~(\ref{srmdcfundef}) in terms of the
 short-range MD correlation integrand of Eq.~(\ref{srmdcintegrand}).
 Since the long-range
 wavefunction in the dissociation limit of H$_2$ reduces to the
 Heitler--London wavefunction for all non-zero $\mu$
 values,\cite{PaolasrXmd} this integrand and, consequently, the
 short-range MD correlation energy should vanish upon bond
 stretching. 
This is an important difference between MC-srDFT and RSMCHf schemes.
While the former is expected to describe much of the
static correlation with the short-range correlation functional,
the latter treats all static correlation with MCSCF, at
least in the dissociation limit.

Comparing the RSMCHf and RSMCHf (no src) PECs of H$_2$ at large
separations, it is clear that the short-range MD correlation energy
is not well described within the local density approximation, as expected from the work of Gori-Giorgi and Savin.\cite{PaolasrXmd} 
The error can be interpreted as a
 self-correlation error of the spin-unpolarized MD srLDA functional since the molecule is correctly dissociated into
 two neutral hydrogen atoms (see the MC-srLDA natural orbital
 occupancies in Fig.~\ref{Occnumberswithdist}). It is, however, not
 exclusively due to the functional---indeed, the RSMCHf (no src) PEC deviates slightly from the exact PEC.
 The deviation comes from the MC-srLDA wavefunction that is used to
 compute the RSMCHf (no src) energy. The former contains self-interaction
 errors because of the srLDA exchange--correlation density-functional
 potential. This problem will be addressed in the following sections.
 
Finally, we observe for H$_2$ a slight bump in the intermediate 
region ($R=2.25$\,\AA) of the RSMCHf PEC. It is even more pronounced for H$_2$O and
N$_2$. No bump appears for Li$_2$, possibly because (unlike
the other systems) it 
has a significant multi-configuration character already at equilibrium
(see Fig.~\ref{Occnumberswithdist}).
Interestingly, at $R=2.117$\,\AA, the RSMCHf (no src)
energy of H$_2$ deviates from the exact one by about 0.03 $E_\text h$,
which could be interpreted as
the correct value for the short-range MD correlation energy (the
RSMCHf energy is relatively close to the exact one). However, from the
srOEP calculations of Gori-Giorgi and Savin 
(the lowest panel in Fig.\;7 of Ref.\;\onlinecite{PaolasrXmd}),
the accurate value of this energy for $\mu=0.4$ is 0.01--0.015\,$E_\text h$.
This difference suggests that
the srLDA exchange--correlation density-functional potential, from which the MC-srLDA
wavefunction (and hence the RSMCHf energy) is obtained, may not be
accurate enough, especially as the wavefunction is strongly
multi-configurational in this region, as shown in Fig.~\ref{Occnumberswithdist}. 
Interestingly, errors on the short-range
potential and the short-range MD correlation density functional 
seem to compensate in the RSMCHf energy when $R=2.117$\,\AA~but not for
the other bond distances. Use of
srOEPs rather than srLDA potentials is then a reasonable alternative, as is investigated in the following.   

\subsubsection{MC-srFEP results}\label{subsubsec:mc-srfep_results}

The MC-srFEP model that was introduced in Sec.~\ref{subsec:summary} corresponds to a zero-iteration
MC-srOEP calculation where the long-range MC wavefunction only is
optimized. The initial HF-srOEP potential is simply frozen. Equilibrium bond distances and binding energies obtained at the MC-srFEP
level are given in Tables \ref{ReDeRe} and \ref{ReDeRe_second_half}. Full PECs have been plotted in
Fig.~\ref{PECandIE_curves} for H$_2$ and Li$_2$. 
Convergence problems in the HF-srOEP calculation occurred for 1.6 $\leq R \leq $ 3.5\,{\AA} in N$_2$ and for all O--H
distances beyond 2.0\,{\AA} in H$_2$O---that is, when the long-range wavefunction becomes strongly 
multi-configurational, as shown in Fig.~\ref{Occnumberswithdist}. In these cases, we expect the convergence
of the srOEP to be manageable only at the long-range MCSCF level, requiring
the implementation of analytical gradients, which is currently in
progress.

For H$_2$ and Li$_2$, we observe that the MC-srFEP and RSMCHf PECs remain relatively close for all
bond distances, the natural orbital occupations being almost identical (see Fig.\;\ref{Occnumberswithdist}). 
For H$_2$ at equilibrium, the MC-srFEP energy is slightly lower than the RSMCHf energy, meaning that
the HF-srOEP potential is better than the srLDA one, according
to the variational principle in Eq.~(\ref{GsenevmuMC}). 
However, as the bond is
stretched and the wavefunction becomes multi-configurational, the MC-srFEP energy becomes higher than the RSMCHf one. This
is expected since the HF-srOEP potential is unaffected by long-range
correlation. 
For comparison, the MC-srFEP (no src) PEC has been computed. In this
case the frozen potential is calculated at the HF-srOEP (no src) level,
that is without the MD srLDA correlation
term. In the dissociation limit, the MC-srFEP (no src) energy is too
high which clearly indicates that the HF-srOEP (no src) potential is a poor approximation to the exact short-range
potential. Optimization of the srOEP at the long-range MCSCF level should, 
on the other hand, give essentially the exact solution.\cite{PaolasrXmd}
It is noteworthy that, in a minimal atomic-orbital basis, the exact
energy would actually be
obtained at both RSMCHf (no src) and MC-srFEP (no src) 
levels of theory since the wavefunction is then fixed to 
$1/\sqrt{2}\left(\vert1\sigma_g^2\rangle-\vert1\sigma_u^2\rangle\right)$,
where the bonding $1\sigma_g$ and the anti-bonding $1\sigma_u$ molecular
orbitals are simply linear combinations of the atomic $1s$ orbitals
of each hydrogen atom.\cite{Sharkas_JCP}
In our calculations, as larger basis are used, orbitals can rotate so
that different energies can be obtained.

Returning to RSMCHf, the srLDA density-functional 
potential, even though it may be a crude approximation to the exact
potential, includes multi-configuration effects through
the density. It is therefore more accurate than the HF-srOEP potential. 
Note also that, in the intermediate region, the
maximum of the bump is at a higher energy with the MC-srFEP model than with the RSMCHf model. 
In conclusion, MC-srFEP provides better binding energies than
RSMCHf but this essentially relies on error compensation. 

\subsubsection{MC-srOEP results for H$_2$}\label{subsubsec:mc-sroep-h2}

We now consider the calculation of the PEC of H$_2$ at the
MC-srOEP level. 
For comparison, the MC-srOEP (no src) PEC was also calculated. These PECs are shown in Fig.~\ref{PECandIE_curves}.
As expected, the MC-srOEP and MC-srFEP PECs are almost identical in the
equilibrium region where static correlation is negligible.
As in the MC-srLDA scheme, the MC-srOEP wavefunction is essentially a single
determinant (see Fig.~\ref{Occnumberswithdist}).

Upon bond stretching, the MC-srOEP PEC deviates
from the MC-srFEP curve. Interestingly, the separation occurs when the
MC-srFEP energy is higher than the RSMCHf one, that is when  
the HF-srOEP potential becomes less accurate than the srLDA one, as discussed
previously. 
The
difference between MC-srOEP and MC-srFEP energies is reflected both in the orbital occupations and in
the orbitals (not shown). The MC-srOEP wavefunction has a much more pronounced
multi-configurational character than the MC-srFEP wavefunction, as
shown in
Fig.\;\ref{Occnumberswithdist}. This explains why the MC-srOEP energy reaches it asymptotic limit at a shorter bond distance
($R\approx2.2$\,\AA) than does the MC-srFEP energy ($R\approx3.5$\,\AA). 
As expected, the exact energy is recovered at the MC-srOEP (no src) level in the dissociation limit.
Due to self-correlation errors in the MD
srLDA functional, the MC-srOEP energy is too low at the dissociation.
This 
leads to an underestimation of the binding energy, which is actually
even more pronounced than for RSMCHf and MC-srFEP.

In the intermediate region, at $R=2.117$\,{\AA},
the MC-srOEP (no src) energy differs from the exact one by 0.017$E_\text h$,
as expected for the accurate short-range MD
correlation energy from the work of Gori-Giorgi and Savin.\cite{PaolasrXmd} 
The MD srLDA correlation energy obtained by us ($-0.03E_\text h$) agrees perfectly with their value, 
see the lowest panel in Fig.\;7 of Ref.\;\onlinecite{PaolasrXmd}.
Note also that the bump observed at the RSMCHf and MC-srFEP levels is significantly reduced at the MC-srOEP level 
but not completely removed.

In conclusion, our preliminary MC-srOEP calculations on H$_2$
confirm that the MD srLDA correlation functional of
Paziani {\it et al.}~\cite{srmdCref} can be improved upon. The conclusion is basically the
same as for MC-srLDA: better short-range functionals should be
developed. However, the exact complementary short-range correlation energy to be modelled (namely the MD one) is supposed to vanish in
the dissociation limit, unlike the KS short-range correlation energy
that is used in
conventional MC-srDFT models. 
It may therefore be easier to develop better short-range
MD correlation density functionals. The {\it ab initio} calculation of the
range-separated adiabatic connection~\cite{AMTOptseplrsr} is a valuable tool for such
developments. 

\subsubsection{Analysis of the srOEPs}

\begin{figure}
{\includegraphics[width=0.7\textwidth]{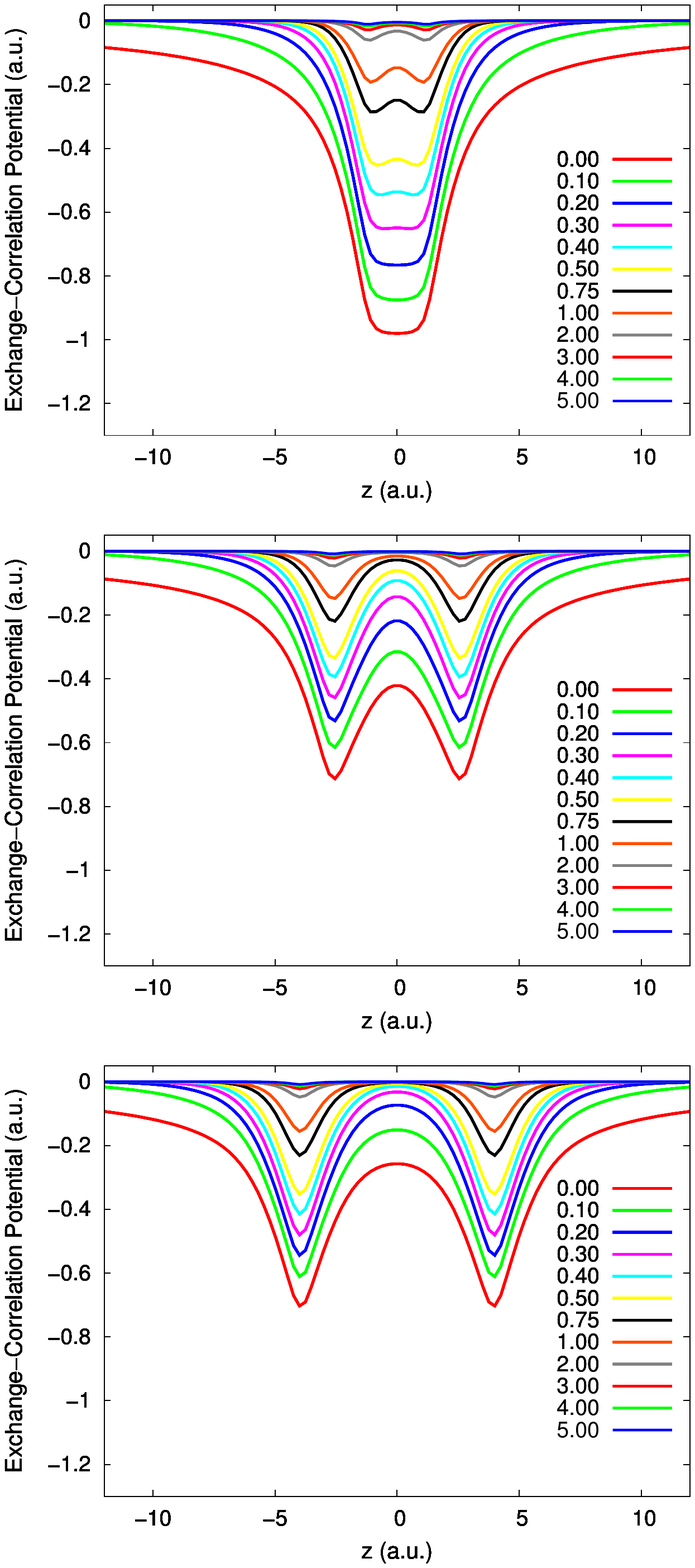}}
\caption{\label{Pots} The exchange--correlation contributions to the srOEPs plotted along the bond axis for the H$_2$ molecule at the HF-srOEP level. The top panel corresponds to $R=1.3$\AA, the middle to $R=2.8$\AA and the bottom to $R=4.2$\AA. In each panel the potentials are shown for $\mu=0.00,0.10,0.20,0.30,0.40,0.50,0.75,1.00,2.00,3.00,4.00,5.00$ and these may be distinguished by noting that the potential value at $z=0.0$ increases with increasing $\mu$.}
\end{figure}

When applying the OEP method care must be taken to ensure that smooth potentials are obtained. In the present work we have used uncontracted basis sets and chosen the orbital and potential sets identical to help ensure this is the case. In Fig.~\ref{Pots} we present the exchange--correlation contributions to the HF-srOEP potentials. For this figure we have generated the potentials using the smoothing-norm approach of Heaton-Burgess \emph{et al.}\cite{Heaton-Burgess1,Heaton-Burgess} with a smoothing parameter of $10^{-5}$. This value perturbs the energy by less than $10^{-5}$a.u. from the unconstrained energies presented in the rest of this work. The result is a a potential that is everywhere smooth. Without the application of this approach the potential very close to the nuclei exhibits a large spike, owing to the fact that this region has essentially no contribution when determining the energy of the system. The potential in other regions of space is essentially unchanged from its unconstrained counterpart.

In the top panel of Fig.~\ref{Pots} the potentials exhibit relatively few features, although as $\mu$ increases the positions of the nuclei can be discerned. As would be expected the from the discussion in Sec.~\ref{THEORY} the srOEP potential approaches zero as the value of $\mu$ is increased. The other feature of the potentials that is clearly visible is their rate of decay as a function of $\mu$. For $\mu=0.00$ a KS-OEP is obtained and exhibits the usual $-1/r$ asymptotic decay. As mu is increased the decay of the potential becomes more rapid, reflecting the form of Eq.~(\ref{vgaussian}). For $\mu=0.4$ the potential contributions along the bond axis are essentially zero beyond approximately $5$ a.u. from each atom. Interestingly, the potential near the atoms is also affected rather strongly by the change in $\mu$, this may again reflect the inclusion of long-range / short-range coupling effects in the partitioning of Toulouse \emph{et al.}~\cite{TousrXmd}. For the longer bond lengths in the lower two panels of Fig.~\ref{Pots} similar conclusions can be reached, though as the atoms are further separated their positions become clearer in the associated potentials. 

For the MC-srOEP approach the exchange--correlation potentials obtained (not shown) are very similar, having only very slightly more negative potentials surrounding the nuclei and slightly more positive potentials further away in the intermediate ``shoulder'' regions. Whilst these subtle changes are essential for proper optimization of the MC-srOEP energy, they do not significantly alter the potentials from those of the HF-srOEP approach on the scale shown.  

\section{Conclusions}\label{conclusion_sec}

An alternative separation of exchange and correlation energies has been
investigated in the context of multi-configuration range-separated DFT.
The new decomposition of the short-range exchange--correlation energy relies
on the auxiliary long-range interacting wavefunction rather than the KS
determinant. 
This approach, first proposed by Toulouse {\it et al.},\cite{TousrXmd}
has two advantages relative to the traditional KS decomposition. First,
the MCSCF part of the energy is now computed 
with the regular (fully-interacting) Hamiltonian, following CAS-DFT
approaches.\cite{DoubleCountothers1,DoubleCountothers2} Second, the
exact complementary short-range
correlation energy vanishes upon dissociation of H$_2$, meaning that the static correlation is fully
assigned to MCSCF. 

The drawback is that, because of double counting, the long-range interacting wave
function used to compute the energy cannot be obtained
by minimizing the energy expression with respect to
the wavefunction parameters. Different approaches that overcome this problem
have been investigated. The first simply computes the energy from
the long-range interacting MCSCF wavefunction that is optimized with the
traditional KS decomposition of the short-range density-functional
energy. A more sophisticated scheme uses short-range
OEPs. The resulting combination of OEP techniques with wavefunction
theory has been investigated in this work, at the HF and MCSCF
levels. 

In the HF case, an analytical expression for the energy
gradient has been derived and implemented. Calculations have been
performed within the short-range local density approximation on H$_2$, N$_2$, Li$_2$ and
H$_2$O. Significant improvements in binding energies are obtained with
the new decomposition of the short-range energy, relative to the traditional one. The importance of
optimizing the short-range OEP at the MCSCF level when static correlation
becomes significant has been demonstrated for
H$_2$, using a finite-difference gradient. 
For further assessment of this approach, the analytical gradient is
under implementation. 

Our preliminary calculations indicate that the local density
approximation is not
accurate enough for modelling the complementary short-range correlation
energy---better functionals may be developed from an accurate
calculation of range-separated adiabatic-connection paths. 
Work is in progress in this direction.   

\section*{Acknowledgments}
E.F. and A.S. thank ANR (DYQUMA project) as well as 
Hans J\o rgen Aa.~Jensen, Andreas Savin and Yann Cornaton for fruitful discussions.
  T. H. and A. M. T.
  acknowledge supported by the Norwegian Research Council through
  the CoE Centre for Theoretical and Computational Chemistry (CTCC)
  Grant No.\ 179568/V30 and by the
  European Research Council under the European Union Seventh Framework
  Program through the Advanced Grant ABACUS, ERC Grant Agreement No.\
  267683. A. M. T. is also grateful for support from the Royal Society University Research Fellowship scheme.
  A. S. acknowledges support from the CTCC for a visit to Oslo.


\begin{thebibliography}{83}
\expandafter\ifx\csname natexlab\endcsname\relax\def\natexlab#1{#1}\fi
\expandafter\ifx\csname bibnamefont\endcsname\relax
  \def\bibnamefont#1{#1}\fi
\expandafter\ifx\csname bibfnamefont\endcsname\relax
  \def\bibfnamefont#1{#1}\fi
\expandafter\ifx\csname citenamefont\endcsname\relax
  \def\citenamefont#1{#1}\fi
\expandafter\ifx\csname url\endcsname\relax
  \def\url#1{\texttt{#1}}\fi
\expandafter\ifx\csname urlprefix\endcsname\relax\def\urlprefix{URL }\fi
\providecommand{\bibinfo}[2]{#2}
\providecommand{\eprint}[2][]{\url{#2}}

\bibitem[{\citenamefont{Kohn and Sham}(1965)}]{kstheo}
\bibinfo{author}{\bibfnamefont{W.}~\bibnamefont{Kohn}} \bibnamefont{and}
  \bibinfo{author}{\bibfnamefont{L.~J.} \bibnamefont{Sham}},
  \bibinfo{journal}{Phys. Rev. A} \textbf{\bibinfo{volume}{140}},
  \bibinfo{pages}{1133} (\bibinfo{year}{1965}).

\bibitem[{\citenamefont{Malet and Gori-Giorgi}(2012)}]{Gori-Giorgi_PRL_2012}
\bibinfo{author}{\bibfnamefont{F.}~\bibnamefont{Malet}} \bibnamefont{and}
  \bibinfo{author}{\bibfnamefont{P.}~\bibnamefont{Gori-Giorgi}},
  \bibinfo{journal}{Phys. Rev. Lett.} \textbf{\bibinfo{volume}{109}},
  \bibinfo{pages}{246402} (\bibinfo{year}{2012}).

\bibitem[{\citenamefont{Becke}(2013)}]{dft-becke_JCP2013}
\bibinfo{author}{\bibfnamefont{A.~D.} \bibnamefont{Becke}},
  \bibinfo{journal}{J. Chem. Phys.} \textbf{\bibinfo{volume}{138}},
  \bibinfo{pages}{074109} (\bibinfo{year}{2013}).

\bibitem[{\citenamefont{Schipper et~al.}(1998)\citenamefont{Schipper,
  Gritsenko, and Baerends}}]{ensemble_Baerends_1998}
\bibinfo{author}{\bibfnamefont{P.~R.~T.} \bibnamefont{Schipper}},
  \bibinfo{author}{\bibfnamefont{O.~V.} \bibnamefont{Gritsenko}},
  \bibnamefont{and} \bibinfo{author}{\bibfnamefont{E.~J.}
  \bibnamefont{Baerends}}, \bibinfo{journal}{Theor. Chem. Acc.}
  \textbf{\bibinfo{volume}{99}}, \bibinfo{pages}{329} (\bibinfo{year}{1998}).

\bibitem[{\citenamefont{Filatov and Shaik}(1999)}]{REKS-Filatov-CPL_1999}
\bibinfo{author}{\bibfnamefont{M.}~\bibnamefont{Filatov}} \bibnamefont{and}
  \bibinfo{author}{\bibfnamefont{S.}~\bibnamefont{Shaik}},
  \bibinfo{journal}{Chem. Phys. Lett.} \textbf{\bibinfo{volume}{304}},
  \bibinfo{pages}{429} (\bibinfo{year}{1999}).

\bibitem[{\citenamefont{Chai}(2012)}]{dft-Chai_JCP2012}
\bibinfo{author}{\bibfnamefont{J.-D.} \bibnamefont{Chai}}, \bibinfo{journal}{J.
  Chem. Phys.} \textbf{\bibinfo{volume}{136}}, \bibinfo{pages}{154104}
  (\bibinfo{year}{2012}).

\bibitem[{\citenamefont{Nygaard and Olsen}(2013)}]{can-ensemble-Olsen_2013}
\bibinfo{author}{\bibfnamefont{C.~R.} \bibnamefont{Nygaard}} \bibnamefont{and}
  \bibinfo{author}{\bibfnamefont{J.}~\bibnamefont{Olsen}}, \bibinfo{journal}{J.
  Chem. Phys.} \textbf{\bibinfo{volume}{138}}, \bibinfo{pages}{094109}
  (\bibinfo{year}{2013}).

\bibitem[{\citenamefont{Colle and Salvetti}(1990)}]{ColleandSalvetti_JCP_1990}
\bibinfo{author}{\bibfnamefont{R.}~\bibnamefont{Colle}} \bibnamefont{and}
  \bibinfo{author}{\bibfnamefont{O.}~\bibnamefont{Salvetti}},
  \bibinfo{journal}{J. Chem. Phys.} \textbf{\bibinfo{volume}{93}},
  \bibinfo{pages}{534} (\bibinfo{year}{1990}).

\bibitem[{\citenamefont{Savin}(1988)}]{savinijqc}
\bibinfo{author}{\bibfnamefont{A.}~\bibnamefont{Savin}}, \bibinfo{journal}{Int.
  J. Quantum Chem.} \textbf{\bibinfo{volume}{34}}, \bibinfo{pages}{59}
  (\bibinfo{year}{1988}).

\bibitem[{\citenamefont{Savin}(1989)}]{savinijqc2}
\bibinfo{author}{\bibfnamefont{A.}~\bibnamefont{Savin}}, \bibinfo{journal}{J.
  Chem. Phys.} \textbf{\bibinfo{volume}{86}}, \bibinfo{pages}{757}
  (\bibinfo{year}{1989}).

\bibitem[{\citenamefont{Miehlich et~al.}(1997)\citenamefont{Miehlich, Stoll,
  and Savin}}]{MiehlichStollSavin}
\bibinfo{author}{\bibfnamefont{B.}~\bibnamefont{Miehlich}},
  \bibinfo{author}{\bibfnamefont{H.}~\bibnamefont{Stoll}}, \bibnamefont{and}
  \bibinfo{author}{\bibfnamefont{A.}~\bibnamefont{Savin}},
  \bibinfo{journal}{Mol. Phys.} \textbf{\bibinfo{volume}{91}},
  \bibinfo{pages}{527} (\bibinfo{year}{1997}).

\bibitem[{\citenamefont{Gr{\"a}ffenstein and
  Cremer}(2000)}]{DoubleCountothers1}
\bibinfo{author}{\bibfnamefont{J.}~\bibnamefont{Gr{\"a}ffenstein}}
  \bibnamefont{and} \bibinfo{author}{\bibfnamefont{D.}~\bibnamefont{Cremer}},
  \bibinfo{journal}{Chem. Phys. Lett.} \textbf{\bibinfo{volume}{316}},
  \bibinfo{pages}{569} (\bibinfo{year}{2000}).

\bibitem[{\citenamefont{Stoll}(2003)}]{Stoll2003141}
\bibinfo{author}{\bibfnamefont{H.}~\bibnamefont{Stoll}},
  \bibinfo{journal}{Chem. Phys. Lett.} \textbf{\bibinfo{volume}{376}},
  \bibinfo{pages}{141} (\bibinfo{year}{2003}).

\bibitem[{\citenamefont{Ying et~al.}(2012)\citenamefont{Ying, Su, Chen, Shaik,
  and Wu}}]{Shaik_JChemTheoryComputation}
\bibinfo{author}{\bibfnamefont{F.}~\bibnamefont{Ying}},
  \bibinfo{author}{\bibfnamefont{P.}~\bibnamefont{Su}},
  \bibinfo{author}{\bibfnamefont{Z.}~\bibnamefont{Chen}},
  \bibinfo{author}{\bibfnamefont{S.}~\bibnamefont{Shaik}}, \bibnamefont{and}
  \bibinfo{author}{\bibfnamefont{W.}~\bibnamefont{Wu}}, \bibinfo{journal}{J.
  Chem. Theory Comput.} \textbf{\bibinfo{volume}{8}}, \bibinfo{pages}{1608}
  (\bibinfo{year}{2012}).

\bibitem[{\citenamefont{Malcolm and McDouall}(1998)}]{Malcolm1998121}
\bibinfo{author}{\bibfnamefont{N.~O.~J.} \bibnamefont{Malcolm}}
  \bibnamefont{and} \bibinfo{author}{\bibfnamefont{J.~J.~W.}
  \bibnamefont{McDouall}}, \bibinfo{journal}{Chem. Phys. Lett.}
  \textbf{\bibinfo{volume}{282}}, \bibinfo{pages}{121} (\bibinfo{year}{1998}).

\bibitem[{\citenamefont{McDouall}(2003)}]{JMcdouall_MP_2003}
\bibinfo{author}{\bibfnamefont{J.~J.~W.} \bibnamefont{McDouall}},
  \bibinfo{journal}{Mol. Phys.} \textbf{\bibinfo{volume}{101}},
  \bibinfo{pages}{361} (\bibinfo{year}{2003}).

\bibitem[{\citenamefont{Gr{\"a}ffenstein and
  Cremer}(2005)}]{DoubleCountothers1b}
\bibinfo{author}{\bibfnamefont{J.}~\bibnamefont{Gr{\"a}ffenstein}}
  \bibnamefont{and} \bibinfo{author}{\bibfnamefont{D.}~\bibnamefont{Cremer}},
  \bibinfo{journal}{Mol. Phys.} \textbf{\bibinfo{volume}{103}},
  \bibinfo{pages}{279} (\bibinfo{year}{2005}).

\bibitem[{\citenamefont{Gusarov et~al.}(2004)\citenamefont{Gusarov, Malmqvist,
  Lindh, and Roos}}]{DoubleCountothers2}
\bibinfo{author}{\bibfnamefont{S.}~\bibnamefont{Gusarov}},
  \bibinfo{author}{\bibfnamefont{P.-{\AA}.} \bibnamefont{Malmqvist}},
  \bibinfo{author}{\bibfnamefont{R.}~\bibnamefont{Lindh}}, \bibnamefont{and}
  \bibinfo{author}{\bibfnamefont{B.~O.} \bibnamefont{Roos}},
  \bibinfo{journal}{Theor. Chem. Acc.} \textbf{\bibinfo{volume}{112}},
  \bibinfo{pages}{84} (\bibinfo{year}{2004}).

\bibitem[{\citenamefont{Nakata et~al.}(2006)\citenamefont{Nakata, Ukai,
  Yamanaka, Takada, and Yamaguchi}}]{NakataUkaietal_IJQC_2006}
\bibinfo{author}{\bibfnamefont{K.}~\bibnamefont{Nakata}},
  \bibinfo{author}{\bibfnamefont{T.}~\bibnamefont{Ukai}},
  \bibinfo{author}{\bibfnamefont{S.}~\bibnamefont{Yamanaka}},
  \bibinfo{author}{\bibfnamefont{T.}~\bibnamefont{Takada}}, \bibnamefont{and}
  \bibinfo{author}{\bibfnamefont{K.}~\bibnamefont{Yamaguchi}},
  \bibinfo{journal}{Int. J. Quantum Chem.} \textbf{\bibinfo{volume}{106}},
  \bibinfo{pages}{3325} (\bibinfo{year}{2006}).

\bibitem[{\citenamefont{Yamanaka et~al.}(2006)\citenamefont{Yamanaka, Nakata,
  Ukai, Takada, and Yamaguchi}}]{YamanakaNakataetal_IJQC_2006}
\bibinfo{author}{\bibfnamefont{S.}~\bibnamefont{Yamanaka}},
  \bibinfo{author}{\bibfnamefont{K.}~\bibnamefont{Nakata}},
  \bibinfo{author}{\bibfnamefont{T.}~\bibnamefont{Ukai}},
  \bibinfo{author}{\bibfnamefont{T.}~\bibnamefont{Takada}}, \bibnamefont{and}
  \bibinfo{author}{\bibfnamefont{K.}~\bibnamefont{Yamaguchi}},
  \bibinfo{journal}{Int. J. Quantum Chem.} \textbf{\bibinfo{volume}{106}},
  \bibinfo{pages}{3312} (\bibinfo{year}{2006}).

\bibitem[{\citenamefont{Ukai et~al.}(2007)\citenamefont{Ukai, Nakata, Yamanaka,
  Takada, and Yamaguchi}}]{UkaiNakataetal_MP_2007}
\bibinfo{author}{\bibfnamefont{T.}~\bibnamefont{Ukai}},
  \bibinfo{author}{\bibfnamefont{K.}~\bibnamefont{Nakata}},
  \bibinfo{author}{\bibfnamefont{S.}~\bibnamefont{Yamanaka}},
  \bibinfo{author}{\bibfnamefont{T.}~\bibnamefont{Takada}}, \bibnamefont{and}
  \bibinfo{author}{\bibfnamefont{K.}~\bibnamefont{Yamaguchi}},
  \bibinfo{journal}{Mol. Phys.} \textbf{\bibinfo{volume}{105}},
  \bibinfo{pages}{2667} (\bibinfo{year}{2007}).

\bibitem[{\citenamefont{P\'erez-Jim\'enez and
  P\'erez-Jord\'a}(2007)}]{PhysRevA.75.012503}
\bibinfo{author}{\bibfnamefont{A.~J.} \bibnamefont{P\'erez-Jim\'enez}}
  \bibnamefont{and} \bibinfo{author}{\bibfnamefont{J.~M.}
  \bibnamefont{P\'erez-Jord\'a}}, \bibinfo{journal}{Phys. Rev. A}
  \textbf{\bibinfo{volume}{75}}, \bibinfo{pages}{012503}
  (\bibinfo{year}{2007}).

\bibitem[{\citenamefont{Weimer et~al.}(2008)\citenamefont{Weimer, {Della Sala},
  and G\"{o}rling}}]{JCP08_Gorling_mcoep}
\bibinfo{author}{\bibfnamefont{M.}~\bibnamefont{Weimer}},
  \bibinfo{author}{\bibfnamefont{F.}~\bibnamefont{{Della Sala}}},
  \bibnamefont{and}
  \bibinfo{author}{\bibfnamefont{A.}~\bibnamefont{G\"{o}rling}},
  \bibinfo{journal}{J. Chem. Phys.} \textbf{\bibinfo{volume}{128}},
  \bibinfo{pages}{144109} (\bibinfo{year}{2008}).

\bibitem[{\citenamefont{Kurzweil et~al.}(2009)\citenamefont{Kurzweil, Lawler,
  and Head-Gordon}}]{MCDFT_Head-Gordon_MolPhys2009}
\bibinfo{author}{\bibfnamefont{Y.}~\bibnamefont{Kurzweil}},
  \bibinfo{author}{\bibfnamefont{K.~V.} \bibnamefont{Lawler}},
  \bibnamefont{and}
  \bibinfo{author}{\bibfnamefont{M.}~\bibnamefont{Head-Gordon}},
  \bibinfo{journal}{Mol. Phys.} \textbf{\bibinfo{volume}{107}},
  \bibinfo{pages}{2103} (\bibinfo{year}{2009}).

\bibitem[{\citenamefont{Savin}(1996)}]{savinbook}
\bibinfo{author}{\bibfnamefont{A.}~\bibnamefont{Savin}},
  \emph{\bibinfo{title}{Recent Developments and Applications of Modern Density
  Functional Theory}} (\bibinfo{publisher}{Elsevier Amsterdam},
  \bibinfo{year}{1996}), p. \bibinfo{pages}{327}.

\bibitem[{\citenamefont{Stoll and Savin}(1985)}]{stollsavinbook}
\bibinfo{author}{\bibfnamefont{H.}~\bibnamefont{Stoll}} \bibnamefont{and}
  \bibinfo{author}{\bibfnamefont{A.}~\bibnamefont{Savin}},
  \emph{\bibinfo{title}{Density functional methods in physics}}
  (\bibinfo{publisher}{Plenum}, \bibinfo{address}{New York},
  \bibinfo{year}{1985}), p. \bibinfo{pages}{177}.

\bibitem[{\citenamefont{Pedersen}(2004)}]{jesperthesis}
\bibinfo{author}{\bibfnamefont{J.~K.} \bibnamefont{Pedersen}}, Ph.D. thesis,
  \bibinfo{school}{University of Southern Denmark} (\bibinfo{year}{2004}).

\bibitem[{\citenamefont{Fromager et~al.}(2007)\citenamefont{Fromager, Toulouse,
  and Jensen}}]{dft-Fromager-JCP2007a}
\bibinfo{author}{\bibfnamefont{E.}~\bibnamefont{Fromager}},
  \bibinfo{author}{\bibfnamefont{J.}~\bibnamefont{Toulouse}}, \bibnamefont{and}
  \bibinfo{author}{\bibfnamefont{H.~J.~{\Aa}.} \bibnamefont{Jensen}},
  \bibinfo{journal}{J. Chem. Phys.} \textbf{\bibinfo{volume}{126}},
  \bibinfo{pages}{074111} (\bibinfo{year}{2007}).

\bibitem[{\citenamefont{Fromager et~al.}(2009)\citenamefont{Fromager, R\'{e}al,
  W{\aa}hlin, Wahlgren, and Jensen}}]{JCPunivmu2}
\bibinfo{author}{\bibfnamefont{E.}~\bibnamefont{Fromager}},
  \bibinfo{author}{\bibfnamefont{F.}~\bibnamefont{R\'{e}al}},
  \bibinfo{author}{\bibfnamefont{P.}~\bibnamefont{W{\aa}hlin}},
  \bibinfo{author}{\bibfnamefont{U.}~\bibnamefont{Wahlgren}}, \bibnamefont{and}
  \bibinfo{author}{\bibfnamefont{H.~J.~{\Aa}.} \bibnamefont{Jensen}},
  \bibinfo{journal}{J. Chem. Phys.} \textbf{\bibinfo{volume}{131}},
  \bibinfo{pages}{054107} (\bibinfo{year}{2009}).

\bibitem[{\citenamefont{\'{A}ngy\'{a}n
  et~al.}(2005)\citenamefont{\'{A}ngy\'{a}n, Gerber, Savin, and
  Toulouse}}]{srdftnancy}
\bibinfo{author}{\bibfnamefont{J.~G.} \bibnamefont{\'{A}ngy\'{a}n}},
  \bibinfo{author}{\bibfnamefont{I.~C.} \bibnamefont{Gerber}},
  \bibinfo{author}{\bibfnamefont{A.}~\bibnamefont{Savin}}, \bibnamefont{and}
  \bibinfo{author}{\bibfnamefont{J.}~\bibnamefont{Toulouse}},
  \bibinfo{journal}{Phys. Rev. A} \textbf{\bibinfo{volume}{72}},
  \bibinfo{pages}{012510} (\bibinfo{year}{2005}).

\bibitem[{\citenamefont{Goll et~al.}(2005)\citenamefont{Goll, Werner, and
  Stoll}}]{ccsrdft}
\bibinfo{author}{\bibfnamefont{E.}~\bibnamefont{Goll}},
  \bibinfo{author}{\bibfnamefont{H.~J.} \bibnamefont{Werner}},
  \bibnamefont{and} \bibinfo{author}{\bibfnamefont{H.}~\bibnamefont{Stoll}},
  \bibinfo{journal}{Phys. Chem. Chem. Phys.} \textbf{\bibinfo{volume}{7}},
  \bibinfo{pages}{3917} (\bibinfo{year}{2005}).

\bibitem[{\citenamefont{Toulouse et~al.}(2009)\citenamefont{Toulouse, Gerber,
  Jansen, Savin, and \'{A}ngy\'{a}n}}]{rpa-srdft_toulouse}
\bibinfo{author}{\bibfnamefont{J.}~\bibnamefont{Toulouse}},
  \bibinfo{author}{\bibfnamefont{I.~C.} \bibnamefont{Gerber}},
  \bibinfo{author}{\bibfnamefont{G.}~\bibnamefont{Jansen}},
  \bibinfo{author}{\bibfnamefont{A.}~\bibnamefont{Savin}}, \bibnamefont{and}
  \bibinfo{author}{\bibfnamefont{J.~G.} \bibnamefont{\'{A}ngy\'{a}n}},
  \bibinfo{journal}{Phys. Rev. Lett.} \textbf{\bibinfo{volume}{102}},
  \bibinfo{pages}{096404} (\bibinfo{year}{2009}).

\bibitem[{\citenamefont{Fromager et~al.}(2010)\citenamefont{Fromager,
  Cimiraglia, and Jensen}}]{nevpt2srdft2}
\bibinfo{author}{\bibfnamefont{E.}~\bibnamefont{Fromager}},
  \bibinfo{author}{\bibfnamefont{R.}~\bibnamefont{Cimiraglia}},
  \bibnamefont{and} \bibinfo{author}{\bibfnamefont{H.~J.~{\Aa}.}
  \bibnamefont{Jensen}}, \bibinfo{journal}{Phys. Rev. A}
  \textbf{\bibinfo{volume}{81}}, \bibinfo{pages}{024502}
  (\bibinfo{year}{2010}).

\bibitem[{\citenamefont{Janesko et~al.}(2009)\citenamefont{Janesko, Henderson,
  and Scuseria}}]{rpa-srdft_scuseria}
\bibinfo{author}{\bibfnamefont{B.~G.} \bibnamefont{Janesko}},
  \bibinfo{author}{\bibfnamefont{T.~M.} \bibnamefont{Henderson}},
  \bibnamefont{and} \bibinfo{author}{\bibfnamefont{G.~E.}
  \bibnamefont{Scuseria}}, \bibinfo{journal}{J. Chem. Phys.}
  \textbf{\bibinfo{volume}{130}}, \bibinfo{pages}{081105}
  (\bibinfo{year}{2009}).

\bibitem[{\citenamefont{Toulouse et~al.}(2011)\citenamefont{Toulouse, Zhu,
  Savin, Jansen, and \'{A}ngy\'{a}n}}]{rpa-srdft_toulouse_2011}
\bibinfo{author}{\bibfnamefont{J.}~\bibnamefont{Toulouse}},
  \bibinfo{author}{\bibfnamefont{W.}~\bibnamefont{Zhu}},
  \bibinfo{author}{\bibfnamefont{A.}~\bibnamefont{Savin}},
  \bibinfo{author}{\bibfnamefont{G.}~\bibnamefont{Jansen}}, \bibnamefont{and}
  \bibinfo{author}{\bibfnamefont{J.~G.} \bibnamefont{\'{A}ngy\'{a}n}},
  \bibinfo{journal}{J. Chem. Phys.} \textbf{\bibinfo{volume}{135}},
  \bibinfo{pages}{084119} (\bibinfo{year}{2011}).

\bibitem[{\citenamefont{Str{\o}msheim et~al.}(2011)\citenamefont{Str{\o}msheim,
  Kumar, Coriani, Sagvolden, Teale, and Helgaker}}]{TealeDisp}
\bibinfo{author}{\bibfnamefont{M.~D.} \bibnamefont{Str{\o}msheim}},
  \bibinfo{author}{\bibfnamefont{N.}~\bibnamefont{Kumar}},
  \bibinfo{author}{\bibfnamefont{S.}~\bibnamefont{Coriani}},
  \bibinfo{author}{\bibfnamefont{E.}~\bibnamefont{Sagvolden}},
  \bibinfo{author}{\bibfnamefont{A.~M.} \bibnamefont{Teale}}, \bibnamefont{and}
  \bibinfo{author}{\bibfnamefont{T.}~\bibnamefont{Helgaker}},
  \bibinfo{journal}{J. Chem. Phys.} \textbf{\bibinfo{volume}{135}},
  \bibinfo{pages}{194109} (\bibinfo{year}{2011}).

\bibitem[{\citenamefont{Teale et~al.}(2010)\citenamefont{Teale, Coriani, and
  Helgaker}}]{AMTOptseplrsr}
\bibinfo{author}{\bibfnamefont{A.~M.} \bibnamefont{Teale}},
  \bibinfo{author}{\bibfnamefont{S.}~\bibnamefont{Coriani}}, \bibnamefont{and}
  \bibinfo{author}{\bibfnamefont{T.}~\bibnamefont{Helgaker}},
  \bibinfo{journal}{J. Chem. Phys.} \textbf{\bibinfo{volume}{133}},
  \bibinfo{pages}{164112} (\bibinfo{year}{2010}).

\bibitem[{\citenamefont{Sharkas et~al.}(2012)\citenamefont{Sharkas, Savin,
  Jensen, and Toulouse}}]{Sharkas_JCP}
\bibinfo{author}{\bibfnamefont{K.}~\bibnamefont{Sharkas}},
  \bibinfo{author}{\bibfnamefont{A.}~\bibnamefont{Savin}},
  \bibinfo{author}{\bibfnamefont{H.~J.~{\Aa}.} \bibnamefont{Jensen}},
  \bibnamefont{and} \bibinfo{author}{\bibfnamefont{J.}~\bibnamefont{Toulouse}},
  \bibinfo{journal}{J. Chem. Phys.} \textbf{\bibinfo{volume}{137}},
  \bibinfo{pages}{044104} (\bibinfo{year}{2012}).

\bibitem[{\citenamefont{Toulouse
  et~al.}(2005{\natexlab{a}})\citenamefont{Toulouse, Gori-Giorgi, and
  Savin}}]{TousrXmd}
\bibinfo{author}{\bibfnamefont{J.}~\bibnamefont{Toulouse}},
  \bibinfo{author}{\bibfnamefont{P.}~\bibnamefont{Gori-Giorgi}},
  \bibnamefont{and} \bibinfo{author}{\bibfnamefont{A.}~\bibnamefont{Savin}},
  \bibinfo{journal}{Theor. Chem. Acc.} \textbf{\bibinfo{volume}{114}},
  \bibinfo{pages}{305} (\bibinfo{year}{2005}{\natexlab{a}}).

\bibitem[{\citenamefont{Savin}(1995)}]{SavinDFT}
\bibinfo{author}{\bibfnamefont{A.}~\bibnamefont{Savin}},
  \emph{\bibinfo{title}{Recent advances in density functional methods}}
  (\bibinfo{publisher}{World Scientific}, \bibinfo{address}{Singapore},
  \bibinfo{year}{1995}), vol.~\bibinfo{volume}{1}, pp.
  \bibinfo{pages}{129--153}.

\bibitem[{\citenamefont{Levy}(1979)}]{LevyF}
\bibinfo{author}{\bibfnamefont{M.}~\bibnamefont{Levy}}, \bibinfo{journal}{Proc.
  Natl. Acad. Sci. USA} \textbf{\bibinfo{volume}{76}}, \bibinfo{pages}{6062}
  (\bibinfo{year}{1979}).

\bibitem[{\citenamefont{Lieb}(1983)}]{LiebF}
\bibinfo{author}{\bibfnamefont{E.~H.} \bibnamefont{Lieb}},
  \bibinfo{journal}{Int. J. Quantum Chem.} \textbf{\bibinfo{volume}{24}},
  \bibinfo{pages}{243} (\bibinfo{year}{1983}).

\bibitem[{\citenamefont{Yang}(1998)}]{Yang:1998p441}
\bibinfo{author}{\bibfnamefont{W.}~\bibnamefont{Yang}}, \bibinfo{journal}{J.
  Chem. Phys.} \textbf{\bibinfo{volume}{109}}, \bibinfo{pages}{10107}
  (\bibinfo{year}{1998}).

\bibitem[{\citenamefont{Savin et~al.}(2003)\citenamefont{Savin, Colonna, and
  Pollet}}]{SavRev}
\bibinfo{author}{\bibfnamefont{A.}~\bibnamefont{Savin}},
  \bibinfo{author}{\bibfnamefont{F.}~\bibnamefont{Colonna}}, \bibnamefont{and}
  \bibinfo{author}{\bibfnamefont{R.}~\bibnamefont{Pollet}},
  \bibinfo{journal}{Int. J. Quantum Chem.} \textbf{\bibinfo{volume}{93}},
  \bibinfo{pages}{166} (\bibinfo{year}{2003}).

\bibitem[{\citenamefont{Hohenberg and Kohn}(1964)}]{hktheo}
\bibinfo{author}{\bibfnamefont{P.}~\bibnamefont{Hohenberg}} \bibnamefont{and}
  \bibinfo{author}{\bibfnamefont{W.}~\bibnamefont{Kohn}},
  \bibinfo{journal}{Phys. Rev.} \textbf{\bibinfo{volume}{136}},
  \bibinfo{pages}{B864} (\bibinfo{year}{1964}).

\bibitem[{\citenamefont{Toulouse
  et~al.}(2004{\natexlab{a}})\citenamefont{Toulouse, Colonna, and
  Savin}}]{erferfgaufunc}
\bibinfo{author}{\bibfnamefont{J.}~\bibnamefont{Toulouse}},
  \bibinfo{author}{\bibfnamefont{F.}~\bibnamefont{Colonna}}, \bibnamefont{and}
  \bibinfo{author}{\bibfnamefont{A.}~\bibnamefont{Savin}},
  \bibinfo{journal}{Phys. Rev. A} \textbf{\bibinfo{volume}{70}},
  \bibinfo{pages}{062505} (\bibinfo{year}{2004}{\natexlab{a}}).

\bibitem[{\citenamefont{Toulouse and Savin}(2006)}]{TouSav_srlrcorr}
\bibinfo{author}{\bibfnamefont{J.}~\bibnamefont{Toulouse}} \bibnamefont{and}
  \bibinfo{author}{\bibfnamefont{A.}~\bibnamefont{Savin}}, \bibinfo{journal}{J.
  Mol. Struct. (THEOCHEM)} \textbf{\bibinfo{volume}{762}}, \bibinfo{pages}{147}
  (\bibinfo{year}{2006}).

\bibitem[{\citenamefont{Toulouse
  et~al.}(2004{\natexlab{b}})\citenamefont{Toulouse, Savin, and Flad}}]{toulda}
\bibinfo{author}{\bibfnamefont{J.}~\bibnamefont{Toulouse}},
  \bibinfo{author}{\bibfnamefont{A.}~\bibnamefont{Savin}}, \bibnamefont{and}
  \bibinfo{author}{\bibfnamefont{H.~J.} \bibnamefont{Flad}},
  \bibinfo{journal}{Int. J. Quantum Chem.} \textbf{\bibinfo{volume}{100}},
  \bibinfo{pages}{1047} (\bibinfo{year}{2004}{\natexlab{b}}).

\bibitem[{\citenamefont{Paziani et~al.}(2006)\citenamefont{Paziani, Moroni,
  Gori-Giorgi, and Bachelet}}]{srmdCref}
\bibinfo{author}{\bibfnamefont{S.}~\bibnamefont{Paziani}},
  \bibinfo{author}{\bibfnamefont{S.}~\bibnamefont{Moroni}},
  \bibinfo{author}{\bibfnamefont{P.}~\bibnamefont{Gori-Giorgi}},
  \bibnamefont{and} \bibinfo{author}{\bibfnamefont{G.~B.}
  \bibnamefont{Bachelet}}, \bibinfo{journal}{Phys. Rev. B}
  \textbf{\bibinfo{volume}{73}}, \bibinfo{pages}{155111}
  (\bibinfo{year}{2006}).

\bibitem[{\citenamefont{Heyd et~al.}(2003)\citenamefont{Heyd, Scuseria, and
  Ernzerhof}}]{pbehsea}
\bibinfo{author}{\bibfnamefont{J.}~\bibnamefont{Heyd}},
  \bibinfo{author}{\bibfnamefont{G.~E.} \bibnamefont{Scuseria}},
  \bibnamefont{and}
  \bibinfo{author}{\bibfnamefont{M.}~\bibnamefont{Ernzerhof}},
  \bibinfo{journal}{J. Chem. Phys.} \textbf{\bibinfo{volume}{118}},
  \bibinfo{pages}{8207} (\bibinfo{year}{2003}).

\bibitem[{\citenamefont{Heyd and Scuseria}(2004)}]{pbehseb}
\bibinfo{author}{\bibfnamefont{J.}~\bibnamefont{Heyd}} \bibnamefont{and}
  \bibinfo{author}{\bibfnamefont{G.~E.} \bibnamefont{Scuseria}},
  \bibinfo{journal}{J. Chem. Phys.} \textbf{\bibinfo{volume}{120}},
  \bibinfo{pages}{7274} (\bibinfo{year}{2004}).

\bibitem[{\citenamefont{Goll et~al.}(2006)\citenamefont{Goll, Werner, Stoll,
  Leininger, Gori-Giorgi, and Savin}}]{ccsrdft2}
\bibinfo{author}{\bibfnamefont{E.}~\bibnamefont{Goll}},
  \bibinfo{author}{\bibfnamefont{H.~J.} \bibnamefont{Werner}},
  \bibinfo{author}{\bibfnamefont{H.}~\bibnamefont{Stoll}},
  \bibinfo{author}{\bibfnamefont{T.}~\bibnamefont{Leininger}},
  \bibinfo{author}{\bibfnamefont{P.}~\bibnamefont{Gori-Giorgi}},
  \bibnamefont{and} \bibinfo{author}{\bibfnamefont{A.}~\bibnamefont{Savin}},
  \bibinfo{journal}{Chem. Phys.} \textbf{\bibinfo{volume}{329}},
  \bibinfo{pages}{276} (\bibinfo{year}{2006}).

\bibitem[{\citenamefont{Toulouse
  et~al.}(2005{\natexlab{b}})\citenamefont{Toulouse, Colonna, and
  Savin}}]{TouColSavin_JCP2005}
\bibinfo{author}{\bibfnamefont{J.}~\bibnamefont{Toulouse}},
  \bibinfo{author}{\bibfnamefont{F.}~\bibnamefont{Colonna}}, \bibnamefont{and}
  \bibinfo{author}{\bibfnamefont{A.}~\bibnamefont{Savin}}, \bibinfo{journal}{J.
  Chem. Phys.} \textbf{\bibinfo{volume}{122}}, \bibinfo{pages}{014110}
  (\bibinfo{year}{2005}{\natexlab{b}}).

\bibitem[{\citenamefont{Goll et~al.}(2009)\citenamefont{Goll, Ernst,
  Moegle-Hofacker, and Stoll}}]{goll_srmeta-gga}
\bibinfo{author}{\bibfnamefont{E.}~\bibnamefont{Goll}},
  \bibinfo{author}{\bibfnamefont{M.}~\bibnamefont{Ernst}},
  \bibinfo{author}{\bibfnamefont{F.}~\bibnamefont{Moegle-Hofacker}},
  \bibnamefont{and} \bibinfo{author}{\bibfnamefont{H.}~\bibnamefont{Stoll}},
  \bibinfo{journal}{J. Chem. Phys.} \textbf{\bibinfo{volume}{130}},
  \bibinfo{pages}{234112} (\bibinfo{year}{2009}).

\bibitem[{\citenamefont{Gori-Giorgi and Savin}(2009)}]{PaolasrXmd}
\bibinfo{author}{\bibfnamefont{P.}~\bibnamefont{Gori-Giorgi}} \bibnamefont{and}
  \bibinfo{author}{\bibfnamefont{A.}~\bibnamefont{Savin}},
  \bibinfo{journal}{Int. J. Quantum Chem.} \textbf{\bibinfo{volume}{109}},
  \bibinfo{pages}{1950} (\bibinfo{year}{2009}).

\bibitem[{\citenamefont{Yang and Wu}(2002)}]{Yang_prl_oep}
\bibinfo{author}{\bibfnamefont{W.}~\bibnamefont{Yang}} \bibnamefont{and}
  \bibinfo{author}{\bibfnamefont{Q.}~\bibnamefont{Wu}}, \bibinfo{journal}{Phys.
  Rev. Lett.} \textbf{\bibinfo{volume}{89}}, \bibinfo{pages}{143002}
  (\bibinfo{year}{2002}).

\bibitem[{\citenamefont{Fromager et~al.}(2013)\citenamefont{Fromager, Knecht,
  and Jensen}}]{fromager2013}
\bibinfo{author}{\bibfnamefont{E.}~\bibnamefont{Fromager}},
  \bibinfo{author}{\bibfnamefont{S.}~\bibnamefont{Knecht}}, \bibnamefont{and}
  \bibinfo{author}{\bibfnamefont{H.~J.~{\Aa}.} \bibnamefont{Jensen}},
  \bibinfo{journal}{J. Chem. Phys.} \textbf{\bibinfo{volume}{138}},
  \bibinfo{pages}{084101} (\bibinfo{year}{2013}).

\bibitem[{\citenamefont{Helgaker et~al.}(2004)\citenamefont{Helgaker,
  J{\o}rgensen, and Olsen}}]{hf_pinkbook}
\bibinfo{author}{\bibfnamefont{T.}~\bibnamefont{Helgaker}},
  \bibinfo{author}{\bibfnamefont{P.}~\bibnamefont{J{\o}rgensen}},
  \bibnamefont{and} \bibinfo{author}{\bibfnamefont{J.}~\bibnamefont{Olsen}},
  \emph{\bibinfo{title}{Molecular Electronic-Structure Theory}}
  (\bibinfo{publisher}{Wiley}, \bibinfo{address}{Chichester},
  \bibinfo{year}{2004}), pp. \bibinfo{pages}{433--522}.

\bibitem[{dal(2011)}]{daltonpack}
\bibinfo{howpublished}{{\sc DALTON2011} an {\it ab initio} electronic structure
  program. See http://daltonprogram.org/} (\bibinfo{year}{2011}).

\bibitem[{\citenamefont{Wu and Yang}(2003)}]{Yang_Jthcompchem_oep}
\bibinfo{author}{\bibfnamefont{Q.}~\bibnamefont{Wu}} \bibnamefont{and}
  \bibinfo{author}{\bibfnamefont{W.}~\bibnamefont{Yang}}, \bibinfo{journal}{J.
  Theor. Comp. Chem.} \textbf{\bibinfo{volume}{2}}, \bibinfo{pages}{627}
  (\bibinfo{year}{2003}).

\bibitem[{\citenamefont{Dunning}(1989)}]{basissets1}
\bibinfo{author}{\bibfnamefont{T.~H.} \bibnamefont{Dunning}},
  \bibinfo{journal}{J. Chem. Phys.} \textbf{\bibinfo{volume}{90}},
  \bibinfo{pages}{1007} (\bibinfo{year}{1989}).

\bibitem[{\citenamefont{Woon and Dunning}(1994)}]{basissets2}
\bibinfo{author}{\bibfnamefont{D.}~\bibnamefont{Woon}} \bibnamefont{and}
  \bibinfo{author}{\bibfnamefont{T.~H.} \bibnamefont{Dunning}},
  \bibinfo{journal}{J. Chem. Phys.} \textbf{\bibinfo{volume}{100}},
  \bibinfo{pages}{2975} (\bibinfo{year}{1994}).

\bibitem[{\citenamefont{Iikura et~al.}(2001)\citenamefont{Iikura, Tsuneda,
  Yanai, and Hirao}}]{hiraomu}
\bibinfo{author}{\bibfnamefont{H.}~\bibnamefont{Iikura}},
  \bibinfo{author}{\bibfnamefont{T.}~\bibnamefont{Tsuneda}},
  \bibinfo{author}{\bibfnamefont{T.}~\bibnamefont{Yanai}}, \bibnamefont{and}
  \bibinfo{author}{\bibfnamefont{K.}~\bibnamefont{Hirao}}, \bibinfo{journal}{J.
  Chem. Phys.} \textbf{\bibinfo{volume}{115}}, \bibinfo{pages}{3540}
  (\bibinfo{year}{2001}).

\bibitem[{\citenamefont{Song et~al.}(2007)\citenamefont{Song, Hirosawa,
  Tsuneda, and Hirao}}]{hiraonewmu}
\bibinfo{author}{\bibfnamefont{J.~W.} \bibnamefont{Song}},
  \bibinfo{author}{\bibfnamefont{T.}~\bibnamefont{Hirosawa}},
  \bibinfo{author}{\bibfnamefont{T.}~\bibnamefont{Tsuneda}}, \bibnamefont{and}
  \bibinfo{author}{\bibfnamefont{K.}~\bibnamefont{Hirao}}, \bibinfo{journal}{J.
  Chem. Phys.} \textbf{\bibinfo{volume}{126}}, \bibinfo{pages}{154105}
  (\bibinfo{year}{2007}).

\bibitem[{\citenamefont{Vydrov et~al.}(2006)\citenamefont{Vydrov, Heyd, Krukau,
  and Scuseria}}]{scuseriacalib1}
\bibinfo{author}{\bibfnamefont{O.~A.} \bibnamefont{Vydrov}},
  \bibinfo{author}{\bibfnamefont{J.}~\bibnamefont{Heyd}},
  \bibinfo{author}{\bibfnamefont{A.~V.} \bibnamefont{Krukau}},
  \bibnamefont{and} \bibinfo{author}{\bibfnamefont{G.~E.}
  \bibnamefont{Scuseria}}, \bibinfo{journal}{J. Chem. Phys.}
  \textbf{\bibinfo{volume}{125}}, \bibinfo{pages}{074106}
  (\bibinfo{year}{2006}).

\bibitem[{\citenamefont{Vydrov and Scuseria}(2006)}]{scuseriacalib2}
\bibinfo{author}{\bibfnamefont{O.~A.} \bibnamefont{Vydrov}} \bibnamefont{and}
  \bibinfo{author}{\bibfnamefont{G.~E.} \bibnamefont{Scuseria}},
  \bibinfo{journal}{J. Chem. Phys.} \textbf{\bibinfo{volume}{125}},
  \bibinfo{pages}{234109} (\bibinfo{year}{2006}).

\bibitem[{\citenamefont{Gerber and \'{A}ngy\'{a}n}(2005)}]{nancycalib}
\bibinfo{author}{\bibfnamefont{I.~C.} \bibnamefont{Gerber}} \bibnamefont{and}
  \bibinfo{author}{\bibfnamefont{J.~G.} \bibnamefont{\'{A}ngy\'{a}n}},
  \bibinfo{journal}{Chem. Phys. Lett.} \textbf{\bibinfo{volume}{415}},
  \bibinfo{pages}{100} (\bibinfo{year}{2005}).

\bibitem[{\citenamefont{Chai and Head-Gordon}(2008)}]{headgordonmu}
\bibinfo{author}{\bibfnamefont{J.~D.} \bibnamefont{Chai}} \bibnamefont{and}
  \bibinfo{author}{\bibfnamefont{M.}~\bibnamefont{Head-Gordon}},
  \bibinfo{journal}{J. Chem. Phys.} \textbf{\bibinfo{volume}{128}},
  \bibinfo{pages}{084106} (\bibinfo{year}{2008}).

\bibitem[{\citenamefont{Rohrdanz et~al.}(2009)\citenamefont{Rohrdanz, Martins,
  and Herbert}}]{dft-Rohrdanz-JCP2009-130-054112}
\bibinfo{author}{\bibfnamefont{M.~A.} \bibnamefont{Rohrdanz}},
  \bibinfo{author}{\bibfnamefont{K.~M.} \bibnamefont{Martins}},
  \bibnamefont{and} \bibinfo{author}{\bibfnamefont{J.~M.}
  \bibnamefont{Herbert}}, \bibinfo{journal}{J. Chem. Phys.}
  \textbf{\bibinfo{volume}{130}}, \bibinfo{pages}{054112}
  (\bibinfo{year}{2009}).

\bibitem[{\citenamefont{Stein et~al.}(2010)\citenamefont{Stein, Eisenberg,
  Kronik, and Baer}}]{PRL10_Baer_tuning_mu_Koopmans}
\bibinfo{author}{\bibfnamefont{T.}~\bibnamefont{Stein}},
  \bibinfo{author}{\bibfnamefont{H.}~\bibnamefont{Eisenberg}},
  \bibinfo{author}{\bibfnamefont{L.}~\bibnamefont{Kronik}}, \bibnamefont{and}
  \bibinfo{author}{\bibfnamefont{R.}~\bibnamefont{Baer}},
  \bibinfo{journal}{Phys. Rev. Lett.} \textbf{\bibinfo{volume}{105}},
  \bibinfo{pages}{266802} (\bibinfo{year}{2010}).

\bibitem[{\citenamefont{Gerber and
  \'{A}ngy\'{a}n}(2007)}]{JCP07_Ian_mp2-srdft_calib_Rg1-Rg2}
\bibinfo{author}{\bibfnamefont{I.~C.} \bibnamefont{Gerber}} \bibnamefont{and}
  \bibinfo{author}{\bibfnamefont{J.~G.} \bibnamefont{\'{A}ngy\'{a}n}},
  \bibinfo{journal}{J. Chem. Phys.} \textbf{\bibinfo{volume}{126}},
  \bibinfo{eid}{044103} (\bibinfo{year}{2007}).

\bibitem[{\citenamefont{Zhu et~al.}(2010)\citenamefont{Zhu, Toulouse, Savin,
  and \'{A}ngy\'{a}n}}]{JCP10_Wuming_rpa-srdft_weak_int}
\bibinfo{author}{\bibfnamefont{W.}~\bibnamefont{Zhu}},
  \bibinfo{author}{\bibfnamefont{J.}~\bibnamefont{Toulouse}},
  \bibinfo{author}{\bibfnamefont{A.}~\bibnamefont{Savin}}, \bibnamefont{and}
  \bibinfo{author}{\bibfnamefont{J.~G.} \bibnamefont{\'{A}ngy\'{a}n}},
  \bibinfo{journal}{J. Chem. Phys.} \textbf{\bibinfo{volume}{132}},
  \bibinfo{eid}{244108} (\bibinfo{year}{2010}).

\bibitem[{\citenamefont{Toulouse et~al.}(2010)\citenamefont{Toulouse, Zhu,
  \'Angy\'an, and Savin}}]{PRA10_Julien_rpa-srDFT}
\bibinfo{author}{\bibfnamefont{J.}~\bibnamefont{Toulouse}},
  \bibinfo{author}{\bibfnamefont{W.}~\bibnamefont{Zhu}},
  \bibinfo{author}{\bibfnamefont{J.~G.} \bibnamefont{\'Angy\'an}},
  \bibnamefont{and} \bibinfo{author}{\bibfnamefont{A.}~\bibnamefont{Savin}},
  \bibinfo{journal}{Phys. Rev. A} \textbf{\bibinfo{volume}{82}},
  \bibinfo{pages}{032502} (\bibinfo{year}{2010}).

\bibitem[{\citenamefont{Lie and Clementi}(1974{\natexlab{a}})}]{lie1}
\bibinfo{author}{\bibfnamefont{G.~C.} \bibnamefont{Lie}} \bibnamefont{and}
  \bibinfo{author}{\bibfnamefont{E.}~\bibnamefont{Clementi}},
  \bibinfo{journal}{J. Chem. Phys.} \textbf{\bibinfo{volume}{60}},
  \bibinfo{pages}{1288} (\bibinfo{year}{1974}{\natexlab{a}}).

\bibitem[{\citenamefont{Lie and Clementi}(1974{\natexlab{b}})}]{lie2}
\bibinfo{author}{\bibfnamefont{G.~C.} \bibnamefont{Lie}} \bibnamefont{and}
  \bibinfo{author}{\bibfnamefont{E.}~\bibnamefont{Clementi}},
  \bibinfo{journal}{J. Chem. Phys.} \textbf{\bibinfo{volume}{60}},
  \bibinfo{pages}{1275} (\bibinfo{year}{1974}{\natexlab{b}}).

\bibitem[{\citenamefont{Coxon and Melville}(2006)}]{liexpdist}
\bibinfo{author}{\bibfnamefont{J.~A.} \bibnamefont{Coxon}} \bibnamefont{and}
  \bibinfo{author}{\bibfnamefont{T.~C.} \bibnamefont{Melville}},
  \bibinfo{journal}{J.~Mol.~Spectroscopy} \textbf{\bibinfo{volume}{235}},
  \bibinfo{pages}{235} (\bibinfo{year}{2006}).

\bibitem[{\citenamefont{Barakat et~al.}(1986)\citenamefont{Barakat, Bacis,
  Carrot, Churassy, Crozet, and Martin}}]{liexpdist2}
\bibinfo{author}{\bibfnamefont{B.}~\bibnamefont{Barakat}},
  \bibinfo{author}{\bibfnamefont{R.}~\bibnamefont{Bacis}},
  \bibinfo{author}{\bibfnamefont{F.}~\bibnamefont{Carrot}},
  \bibinfo{author}{\bibfnamefont{S.}~\bibnamefont{Churassy}},
  \bibinfo{author}{\bibfnamefont{P.}~\bibnamefont{Crozet}}, \bibnamefont{and}
  \bibinfo{author}{\bibfnamefont{F.}~\bibnamefont{Martin}},
  \bibinfo{journal}{Chem.~Phys.} \textbf{\bibinfo{volume}{102}},
  \bibinfo{pages}{215} (\bibinfo{year}{1986}).

\bibitem[{\citenamefont{Cornaton et~al.}(2013)\citenamefont{Cornaton,
  Stoyanova, Jensen, and Fromager}}]{Doublehybmp2_yann}
\bibinfo{author}{\bibfnamefont{Y.}~\bibnamefont{Cornaton}},
  \bibinfo{author}{\bibfnamefont{A.}~\bibnamefont{Stoyanova}},
  \bibinfo{author}{\bibfnamefont{H.~J.~{\Aa}.} \bibnamefont{Jensen}},
  \bibnamefont{and} \bibinfo{author}{\bibfnamefont{E.}~\bibnamefont{Fromager}},
  \bibinfo{journal}{Phys. Rev. A} \textbf{\bibinfo{volume}{88}},
  \bibinfo{pages}{022516} (\bibinfo{year}{2013}).

\bibitem[{\citenamefont{Hasted}(1972)}]{h2oexper}
\bibinfo{author}{\bibfnamefont{J.~B.} \bibnamefont{Hasted}},
  \emph{\bibinfo{title}{Water: A Comprehensive Treatise}}
  (\bibinfo{publisher}{Plenium}, \bibinfo{address}{New York},
  \bibinfo{year}{1972}), vol.~\bibinfo{volume}{1}, p. \bibinfo{pages}{255}.

\bibitem[{\citenamefont{Heaton-Burgess
  et~al.}(2007)\citenamefont{Heaton-Burgess, Bulat, and
  Yang}}]{Heaton-Burgess1}
\bibinfo{author}{\bibfnamefont{T.}~\bibnamefont{Heaton-Burgess}},
  \bibinfo{author}{\bibfnamefont{F.~A.} \bibnamefont{Bulat}}, \bibnamefont{and}
  \bibinfo{author}{\bibfnamefont{W.}~\bibnamefont{Yang}},
  \bibinfo{journal}{Phys. Rev. Lett.} \textbf{\bibinfo{volume}{98}},
  \bibinfo{pages}{256401} (\bibinfo{year}{2007}).

\bibitem[{\citenamefont{Heaton-Burgess and Yang}(2008)}]{Heaton-Burgess}
\bibinfo{author}{\bibfnamefont{T.}~\bibnamefont{Heaton-Burgess}}
  \bibnamefont{and} \bibinfo{author}{\bibfnamefont{W.}~\bibnamefont{Yang}},
  \bibinfo{journal}{J. Chem. Phys.} \textbf{\bibinfo{volume}{129}},
  \bibinfo{pages}{194102} (\bibinfo{year}{2008}).

\bibitem[{\citenamefont{Salek et~al.}(2005)\citenamefont{Salek, Helgaker, and
  Saue}}]{CP05_Pawel_tddft}
\bibinfo{author}{\bibfnamefont{P.}~\bibnamefont{Salek}},
  \bibinfo{author}{\bibfnamefont{T.}~\bibnamefont{Helgaker}}, \bibnamefont{and}
  \bibinfo{author}{\bibfnamefont{T.}~\bibnamefont{Saue}},
  \bibinfo{journal}{Chem. Phys.} \textbf{\bibinfo{volume}{311}},
  \bibinfo{pages}{187 } (\bibinfo{year}{2005}).

\bibitem[{\citenamefont{Perdew et~al.}(1996)\citenamefont{Perdew, Burke, and
  Ernzerhof}}]{h2oexper_de}
\bibinfo{author}{\bibfnamefont{J.~P.} \bibnamefont{Perdew}},
  \bibinfo{author}{\bibfnamefont{K.}~\bibnamefont{Burke}}, \bibnamefont{and}
  \bibinfo{author}{\bibfnamefont{M.}~\bibnamefont{Ernzerhof}},
  \bibinfo{journal}{Phys. Rev. Lett.} \textbf{\bibinfo{volume}{77}},
  \bibinfo{pages}{3865} (\bibinfo{year}{1996}).

\end{thebibliography}

\newcommand{\Aa}[0]{Aa}


\appendix

\section*{Appendix: HF-srOEP linear response equation}

The response of the HF-srOEP determinant related to variations
$\epsilon_t$ in the
srOEP coefficients is obtained when considering the auxiliary energy
\begin{align}\label{fictitiousener}
\mathcal{E}^{\mu}({\boldsymbol \kappa},\epsilon_t)&= \langle \Phi({\boldsymbol \kappa}) \vert\hat{T}+\hat{W}^{\rm lr,\mu}_{\rm ee}+\hat{V_0}\vert \Phi({\boldsymbol \kappa}) \rangle \nonumber\\
&+\sum_t\epsilon_t\langle\Phi({\boldsymbol \kappa})\vert\hat{g}_t\vert\Phi({\boldsymbol \kappa})\rangle \nonumber \\
&=\mathcal{E}^{\mu}({\boldsymbol \kappa})+ \sum_t\epsilon_t \langle\Phi({\boldsymbol \kappa})\vert\hat{g}_t\vert\Phi({\boldsymbol \kappa})\rangle, 
\end{align}
where $\hat{V_0}=
\int {\rm d}{\mathbf r}\;v_0({\mathbf r})\,\hat{n}({\bf r})
$ is the trial srOEP and the Gaussian operator $\hat{g}_t=\int {\rm d}{\mathbf
r}\;g_t({\mathbf r})\,\hat{n}({\bf r})$ is analogous to a property operator 
in response theory~\cite{CP05_Pawel_tddft}. From the variational condition in
Eq.~(\ref{phimuvfromminhf}) we obtain     
\begin{align}\label{fictitiousenerbetaderiv}
 \forall \epsilon_t\hspace{0.3cm}{\left. \frac{\partial\mathcal{E}^{\mu}({\boldsymbol \kappa},\epsilon_t)}{\partial{\boldsymbol \kappa}}\right |_{{\boldsymbol  \kappa}(\epsilon_t)}=0}, 
\end{align}
which leads to the linear response equation
\begin{align}\label{hf-sroep-lreq}
 \left. \frac{\rm d}{\rm d\epsilon_t}\left(\frac{\partial\mathcal{E}^{\mu}({\boldsymbol \kappa},\epsilon_t)}{\partial{\boldsymbol \kappa}}\right)\right |_{0}=0. 
\end{align}
Using the Taylor expansion through second order
\begin{align}\label{Taylorexp}
\mathcal{E}^{\mu}({\boldsymbol \kappa})=\mathcal{E}^{\mu}(0)+ {\frac{1}{2} {\boldsymbol \kappa}^{\rm T}\mathcal{E}^{[2]\mu} \hspace{0.1cm}{\boldsymbol \kappa}+\ldots,} 
\end{align}
where the HF-type Hessian $\mathcal{E}^{[2]\mu}$ is constructed from the auxiliary long-range
interacting Hamiltonian $\hat{T}+\hat{W}^{\rm lr,\mu}_{\rm
ee}+\hat{V_0}$, and rewriting 
the first-order derivative of the Gaussian expectation value
as~\cite{hf_pinkbook} 
\begin{align}
\left.\frac{\partial }{\partial {\boldsymbol  \kappa}}\langle\Phi({\boldsymbol \kappa})\vert\hat{g}_t\vert\Phi({\boldsymbol \kappa})\rangle \right |_{0}=g_t^{[1]},
\end{align}
where the gradient Gaussian vector expression is given in
Eq.~(\ref{gtpropvec}), we finally obtain Eq.~(\ref{linearrspeqfinal}). 


\clearpage



\textbf{TABLE CAPTION}
 
\begin{description}
\item[Table \ref{summary_methods}] 
Wavefunction, local potential and energy expressions associated with all
range-separated methods discussed in this work. Single
determinantal trial wavefunctions are denoted $\Phi$. $\hat{H}=
 \hat{T}+\hat{W}_{\rm ee}+\hat{V}_{\rm ne}
$ and $\hat{H}^{\rm lr,\mu}=\hat{T}+\hat{W}^{\rm lr,\mu}_{\rm
 ee}+\hat{V}_{\rm ne}$ correspond to the physical (fully interacting) and
 long-range interacting Hamiltonians, respectively. $S_M$ denotes the
 active space used in the long-range MCSCF calculation. 
\item[Table \ref{ReDeRe}] 
Equilibrium bond distances R$_{e}$ (\AA) and binding energies D$_{e}$
(eV) for the ground state of H$_2$ and N$_2$. The $\mu$ parameter was set to 0.4. See text for further details.\\
\item[Table \ref{ReDeRe_second_half}] 
Equilibrium bond distances R$_{e}$ (\AA) and binding energies D$_{e}$ (eV) for the ground state of Li$_2$ and of H$_2$O at fixed H-O-H angle of 104.5$^{\circ}$. The $\mu$ parameter was set to 0.4. See text for further details.\\
\end{description}


\clearpage


\textbf{FIGURE CAPTIONS}

\begin{description}

\item [Figure \ref{deltaEmu_curves_all}]
The quantity $\Delta E_{\rm c}^\mu$ for (from top to bottom) H$_2$, N$_2$ and Li$_2$ for each method as a
function of the parameter $\mu$. See text for further details.
\\

\item [Figure \ref{Occnwithmu_curves}]
Occupation numbers of MC-srLDA (dotted), MC-srFEP (dashed) and MC-srOEP (solid)
active natural orbitals as a function of $\mu$ for H$_2$, N$_2$ and Li$_2$ at their experimental equilibrium geometries~\cite{lie1,lie2,liexpdist,liexpdist2}.\\

\item[Figure \ref{PECandIE_curves}]
Potential energy curves (a.~u.) of H$_2$ (upper left panel) and
interaction (binding) energies of N$_2$ (upper right panel), H$_2$O
(lower left panel) and Li$_2$ (lower right panel) obtained by means of
the new multi-configuration range-separated schemes. The range
separation parameter is $\mu$=0.4. The exact interaction energy and PEC curves are taken from Ref.~\onlinecite{lie2} for H$_2$ and Ref.~\onlinecite{lie1} for N$_2$.\\

\item [Figure \ref{Occnumberswithdist}]
Occupation numbers of MC-srLDA (dashed), MC-srFEP (dotted) and MC-srOEP (solid) active natural orbitals as a function of bond distance (\AA) (from top to bottom) H$_2$, Li$_2$ and N$_2$ ~\cite{lie1,lie2,liexpdist,liexpdist2}. The parameter $\mu$ is set to 0.4.\\

\item [Figure \ref{Pots}]
The exchange--correlation contributions to the srOEPs plotted along the
bond axis for the H$_2$ molecule at the HF-srOEP level. The top panel
corresponds to $R=1.3$\AA, the middle to $R=2.8$\AA~and the bottom to $R=4.2$\AA. In each panel the potentials are shown for $\mu=0.00,0.10,0.20,0.30,0.40,0.50,0.75,1.00,2.00,3.00,4.00,5.00$ and these may be distinguished by noting that the potential value at $z=0.0$ increases with increasing $\mu$.

\end{description}

%



\clearpage
\begin{table}
\centering 
\caption{\label{summary_methods} Stoyanova {\it et al.}, Journal of Chemical Physics} 
\scalebox{0.7}{
\begin{tabular}{lccccccc}
\hline \hline
  method & & & wavefunction & & local potential & &  energy
  expression\\
\hline
HF-srLDA & & &
$
\Phi^{\mu}=\!\argmin\limits_\Phi \!\Big\{\!\langle \Phi \vert
\hat{H}^{\rm lr,\mu}
\vert\Phi\rangle 
$
& &
$
v_{\rm ne}({\bf r})+\delta E^{\rm sr,\mu}_{\rm Hxc}/\delta n({\bf r})[n_{\Phi^{\mu}}]$
& & 
$
 \langle \Phi^{\mu} \vert 
 \hat{H}^{\rm lr,\mu}
 \vert \Phi^{\mu} \rangle+E^{\rm sr,\mu}_{\rm
 Hxc}[n_{\Phi^{\mu}}]
$
\\
& & &
$
+
E^{\rm sr,\mu}_{\rm Hxc}[n_{\Phi}]\Big\} 
$
& &
& & 
\\
&&&&&&&\\ 
MC-srLDA & & &
$
\Psi_{M}^{\mu}=\!\argmin\limits_{\Psi\in S_M} \!\Big\{\!\langle \Psi \vert
\hat{H}^{\rm lr,\mu}
\vert\Psi\rangle
$
& &
$
v_{\rm ne}({\bf r})+\delta E^{\rm sr,\mu}_{\rm Hxc}/\delta n({\bf r})[n_{\Psi_M^{\mu}}]$
& & 
$
 \langle \Psi_M^{\mu} \vert 
 \hat{H}^{\rm lr,\mu}
 \vert \Psi_M^{\mu} \rangle+E^{\rm sr,\mu}_{\rm
 Hxc}[n_{\Psi_M^{\mu}}]
$
\\
& & &
$
+
E^{\rm sr,\mu}_{\rm Hxc}[n_{\Psi}] \Big\}
$
& &
& & 
\\
&&&&&&&\\ 
RSHf & & &
$
\Phi^{\mu}
$ [HF-srLDA]
& &
$
v_{\rm ne}({\bf r})+\delta E^{\rm sr,\mu}_{\rm Hxc}/\delta n({\bf r})[n_{\Phi^{\mu}}]$
& & 
$
 \langle \Phi^{\mu} \vert 
 \hat{H}
 \vert \Phi^{\mu} \rangle+E^{\rm sr,\mu}_{\rm
 c, md}[n_{\Phi^{\mu}}]
$
\\
&&&&&&&\\ 
RSMCHf & & &
$
\Psi_{M}^{\mu}
$ [MC-srLDA]
& &
$
v_{\rm ne}({\bf r})+\delta E^{\rm sr,\mu}_{\rm Hxc}/\delta n({\bf r})[n_{\Psi_M^{\mu}}]$
& & 
$
 \langle \Psi_M^{\mu} \vert 
 \hat{H}
 \vert \Psi_M^{\mu} \rangle
 +E^{\rm sr,\mu}_{\rm
 c, md}[n_{\Psi_M^{\mu}}]
$
\\
&&&&&&&\\ 
RSMCHf (no src)& & &
$
\Psi_{M}^{\mu}
$ [MC-srLDA]
& &
$
v_{\rm ne}({\bf r})+\delta E^{\rm sr,\mu}_{\rm Hxc}/\delta n({\bf r})[n_{\Psi_M^{\mu}}]$
& & 
$
 \langle \Psi_M^{\mu} \vert 
 \hat{H}
 \vert \Psi_M^{\mu} \rangle
$
\\
&&&&&&&\\ 
HF-srOEP & & &
$
\Phi^\mu[v_1^{\mu}]
$
& &
$
v_1^{\mu}=\!\argmin\limits_v \!\Big\{\!
\langle \Phi^{\mu}[v] \vert 
\hat{H}
\vert\Phi^{\mu}[v]\rangle
$
& & 
$
 \langle \Phi^{\mu}[v_1^{\mu}] \vert 
 \hat{H}
 \vert \Phi^{\mu}[v_1^{\mu}] \rangle
 +E^{\rm sr,\mu}_{\rm c,
 md}[n_{\Phi^{\mu}[v_1^{\mu}]}]
$
\\
 & & &
& &
$
+ E^{\rm sr,\mu}_{\rm c, md}[n_{\Phi^{\mu}[v]}] 
\Big\},$
& & 
\\
 & & &
& &
$\Phi^{\mu}[v] =\!\argmin\limits_\Phi \!\Big\{\!\langle \Phi \vert \hat{T}+\hat{W}^{\rm lr,\mu}_{\rm ee}\vert\Phi\rangle 
$
& & 
\\
 & & &
& &
$
+ \!\! \int \!\! {\rm d}{\mathbf r}\,v({\mathbf r}){n}_{\Phi}({\bf r})\!
\Big\}
$
& & 
\\
&&&&&&&\\ 
HF-srOEP (no src)& & &
$
\Phi^\mu[v_2^{\mu}]
$
& &
$
v_2^{\mu}=\!\argmin\limits_v \!\Big\{\!
\langle \Phi^{\mu}[v] \vert 
\hat{H}
\vert\Phi^{\mu}[v]\rangle
\Big\}
$
& & 
$
 \langle \Phi^{\mu}[v_2^{\mu}] \vert 
 \hat{H}
 \vert \Phi^{\mu}[v_2^{\mu}] \rangle
$
\\
&&&&&&&\\ 
MC-srFEP & & &
$
\Psi_M^{\mu}[v_1^{\mu}] =\!\argmin\limits_{\Psi\in S_M} \!\Big\{\!\langle
\Psi \vert \hat{T}+\hat{W}^{\rm lr,\mu}_{\rm ee}\vert\Psi\rangle 
$
& &
$
v_1^{\mu}
$ [HF-srOEP]
& & 
$
 \langle \Psi_M^{\mu}[v_1^{\mu}] \vert 
 \hat{H}
 \vert \Psi_M^{\mu}[v_1^{\mu}] \rangle
 +E^{\rm sr,\mu}_{\rm c,
 md}[n_{\Psi_M^{\mu}[v_1^{\mu}]}]
$
\\
 & & &
$+ \!\! \int \!\! {\rm d}{\mathbf r}\,v^\mu_1({\mathbf r}){n}_{\Psi}({\bf r})\!
\Big\}$
& &
& & 
\\
&&&&&&&\\ 
MC-srFEP (no src)& & &
$
\Psi_M^{\mu}[v_2^{\mu}] =\!\argmin\limits_{\Psi\in S_M} \!\Big\{\!\langle
\Psi \vert \hat{T}+\hat{W}^{\rm lr,\mu}_{\rm ee}\vert\Psi\rangle 
$
& &
$
v_2^{\mu}
$ [HF-srOEP (no src)]
& & 
$
 \langle \Psi_M^{\mu}[v_2^{\mu}] \vert 
 \hat{H}
 \vert \Psi_M^{\mu}[v_2^{\mu}] \rangle
$
\\
 & & &
$+ \!\! \int \!\! {\rm d}{\mathbf r}\,v^\mu_2({\mathbf r}){n}_{\Psi}({\bf r})\!
\Big\}$
& &
& & 
\\
&&&&&&&\\ 
MC-srOEP & & &
$
\Psi_M^\mu[v_3^{\mu}]
$
& &
$
v_3^{\mu}=\!\argmin\limits_v \!\Big\{\!
\langle \Psi_M^{\mu}[v] \vert 
\hat{H}
\vert\Psi_M^{\mu}[v]\rangle
$
& & 
$
 \langle \Psi_M^{\mu}[v_3^{\mu}] \vert 
 \hat{H}
 \vert \Psi_M^{\mu}[v_3^{\mu}] \rangle
 +E^{\rm sr,\mu}_{\rm c,
 md}[n_{\Psi_M^{\mu}[v_3^{\mu}]}]
$
\\
 & & &
& &
$
+ E^{\rm sr,\mu}_{\rm c, md}[n_{\Psi_M^{\mu}[v]}] 
\Big\},$
& & 
\\
 & & &
& &
$\Psi_M^{\mu}[v] =\!\argmin\limits_{\Psi\in S_M} \!\Big\{\!\langle \Psi \vert \hat{T}+\hat{W}^{\rm lr,\mu}_{\rm ee}\vert\Psi\rangle 
$
& & 
\\
 & & &
& &
$
+ \!\! \int \!\! {\rm d}{\mathbf r}\,v({\mathbf r}){n}_{\Psi}({\bf r})\!
\Big\}
$
& & 
\\
&&&&&&&\\ 
MC-srOEP (no src)& & &
$
\Psi_M^\mu[v_4^{\mu}]
$
& &
$
v_4^{\mu}=\!\argmin\limits_v \!\Big\{\!
\langle \Psi_M^{\mu}[v] \vert 
\hat{H}
\vert\Psi_M^{\mu}[v]\rangle
\Big\}
$
& & 
$
 \langle \Psi_M^{\mu}[v_4^{\mu}] \vert 
 \hat{H}
 \vert \Psi_M^{\mu}[v_4^{\mu}] \rangle
$
\\
\hline
\hline
 \end{tabular}
 }
\end{table}


\clearpage
\begin{table}
\centering 
\caption{\label{ReDeRe} Stoyanova {\it et al.}, Journal of Chemical Physics} 
\scalebox{0.8}{
\begin{tabular}{lrr}
\hline \hline
   &R$_{e}$&D$_{e}$ \\
\hline
\multicolumn{3}{c}{H$_2$}  \\
HF-srLDA&0.755&10.39\\
RSHf&0.717&11.73 \\
HF-srOEP &0.715&11.75\\
MCSCF&0.755&4.14\\
MC-srLDA&0.756&6.05\\
RSMCHf&0.724&4.41\\
RSMCHf (no src) &0.744&3.91 \\
MC-srFEP&0.720&4.53\\ 
MC-srFEP~(no src)&0.743&4.09\\
MC-srOEP&0.722&4.18\\
MC-srOEP~(no src) &0.743&3.77\\
Exp. &0.741$^{a}$&4.75$^{a}$ \\ \hline 
\multicolumn{3}{c}{N$_2$} \\
HF-srLDA&1.082&30.85 \\
RSHf&1.051&36.86\\
HF-srOEP &1.052&34.55\\
MCSCF&1.105&9.19 \\
MC-srLDA&1.087&16.18 \\
RSMCHf&1.079& 7.96\\
RSMCHf (no src) &1.096&6.86 \\
MC-srFEP &1.077&8.34\\
MC-srFEP~(no src) &1.095& 7.44\\ 
Exp. &1.097$^{a}$&9.91$^{a}$ \\ \hline
 \end{tabular}
 } \\
 \hspace{0pt}$^{a}$Ref.~\onlinecite{lie1}
\end{table}

\clearpage
\begin{table}
\centering 
\caption{\label{ReDeRe_second_half} Stoyanova {\it et al.}, Journal of Chemical Physics} 
\scalebox{0.8}{
\begin{tabular}{lrr}
\hline \hline
   &R$_{e}$&D$_{e}$ \\
\hline
\multicolumn{3}{c}{Li$_2$}  \\
HF-srLDA&2.671&3.273 \\
RSHf&2.722&3.130\\
HF-srOEP&2.722&3.126\\
MCSCF&2.720$^{b}$&1.029$^{b}$ \\
MC-srLDA&2.679&1.091\\
RSMCHf&2.669&1.039\\
RSMCHf~(no src) &2.750& 0.936\\
MC-srFEP &2.666&1.047\\
MC-srFEP~(no src) &2.735&0.951 \\ 
Exp. &2.673$^c$&1.056$^c$\\ \hline
\multicolumn{3}{c}{H$_2$O} \\
HF-srLDA&0.961&19.29 \\
RSHf&0.925&19.30\\
HF-srOEP &0.925&-\\
MCSCF&0.963&8.31\\
MC-srLDA&0.962&13.32 \\
RSMCHf&0.926&9.40 \\
RSMCHf (no src) &0.942&8.10 \\
MC-srFEP &0.927&-\\
MC-srFEP~(no src) &0.944&-\\
 Exp. &0.957$^{d}$&10.06$^{e}$ \\ \hline \hline
 \end{tabular}
 } \\
 \hspace{0pt}$^{b}$NEVPT2: $R_{e}$ =2.720 \AA, \ $D_{e}$=1.043 eV, $^{c}$Ref.~\onlinecite{liexpdist, liexpdist2}, $^{d}$Ref.~\onlinecite{h2oexper}, $^{e}$Ref.~\onlinecite{h2oexper_de}
\end{table}


\end{document}